\documentclass{aa}

\usepackage[varg]{txfonts}
\usepackage{graphicx}
\usepackage{array, booktabs, makecell}
\usepackage{siunitx}
\usepackage[T1]{fontenc}
\usepackage{newtxtext,newtxmath}
\usepackage[version=4]{mhchem}
\usepackage{textcomp, gensymb}
\usepackage[normalem]{ulem}
\usepackage{amsmath}
\usepackage{balance}
\usepackage{tabularx}
\usepackage{amsfonts}
\usepackage{comment}
\usepackage{multicol}
\usepackage[utf8]{inputenc}
\usepackage[english]{babel}
\usepackage{natbib}
\usepackage{subcaption}
\usepackage{geometry}
\usepackage{booktabs}
\usepackage[hidelinks,colorlinks=true,linkcolor=blue,citecolor=blue]{hyperref}
\usepackage[dvipsnames]{xcolor}
\usepackage{longtable}

\usepackage{orcidlink}
\bibpunct{(}{)}{;}{a}{}{,}

\newcommand{\un}[1]{\ensuremath{\mathrm{\,#1}}}

\NewDocumentCommand{\sotwo}{O{red}O{black}+m}
    {%
        \begingroup
        \setulcolor{#1}%
        \setul{-.5ex}{.4pt}%
        \def\SOUL@uleverysyllable{%
            \rlap{%
                \color{#2}\the\SOUL@syllable
                \SOUL@setkern\SOUL@charkern}%
            \SOUL@ulunderline{%
                \phantom{\the\SOUL@syllable}}%
        }%
        \ul{#3}%
        \endgroup
    }

\begin{document}

\title{
Multifrequency simultaneous VLBA view of the radio source 3C\,111} 

\author{V. Bartolini\orcidlink{0009-0008 4659-2917} \inst{1,2}, 
        D. Dallacasa\orcidlink{0000-0003-1246-6492} \inst{2},
        J. L. Gómez\orcidlink{0000-0003-4190-7613} \inst{3},
        M. Giroletti\orcidlink{0000-0002-8657-8852} \inst{2,4}, 
        R. Lico\orcidlink{0000-0001-7361-2460} \inst{4,3}, and
        J. D. Livingston\orcidlink{0000-0002-4090-8000} \inst{1}
     } 
\institute{
\inst{1} Max-Planck-Institut f\"{u}r Radioastronomie, Auf dem H\"{u}gel 69, D-53121 Bonn, Germany \\
\inst{2} Dipartimento di Fisica e Astronomia, Università degli Studi di Bologna, Via Gobetti 93/2, I-40129 Bologna, Italy \\
\inst{3} Instituto de Astrofìsica de Andalucìa-CSIC, Glorieta de la Astronomìa s/n, E-18008 Granada, Spain \\
\inst{4} INAF – Istituto di Radioastronomia, Via Gobetti 101, I-40129 Bologna, Italy \\
}

\date{Received xx / Accepted yy}

  \abstract
{Relativistic jets originating at the center of active galactic nuclei (AGN) are embedded in extreme environments with strong magnetic fields and high particle densities, which makes them a fundamental tool for studying the physics of magnetized plasmas.}
{We aim to investigate the magnetic field structure and the parsec/sub-parsec properties of the relativistic jet in the radio galaxy 3C\,111. Rotation Measure studies of nearby radio-galaxies, such as this one, provide a valuable tool to investigate the transversal magnetic field properties.}
{We use multifrequency simultaneous Very Long Baseline Array (VLBA) data from $5 \un{GHz}$ up to $87.6 \un{GHz}$. We perform an analysis of both total intensity and polarization maps to study the jet magnetic field and infer the spectral properties of the synchrotron emission. We model the brightness distribution of the source with multiple 2D Gaussian components to characterize individual emission features.}
{After determining the core shift $(r_c \propto \nu^{-1.27 \pm 0.19} )$,  we compute the spectral index maps for all the adjacent frequency pairs and find different distributions for the core region ($\alpha_\mathrm{max} \approx 1.5$ and $\alpha_\mathrm{min} \approx 0.2$) and the jet ($\alpha \approx -1.5$ on average) with an unusual optically thick/flat feature in it at $\approx 1-2$ pc from the core. 
Using \texttt{modelfit}, we find a total of 56 components at different frequencies. 
By putting constraints on the size and position, we identify 22 components at different frequencies for which we compute the equipartition magnetic field. 
We compute the rotation measure at two different triplets of frequencies. At $15.2-21.9-43.8\un{GHz} $, we discover high values of $RM$ in the same region where the optically thick/flat feature was found. This can be associated with a high value of electron density at $\approx 1-2$ pc from the core that we interpreted as originating in a cloud of the clumpy torus. At $5-8.4-15.2 \un{GHz} $, we find a distribution of the Electric Vector Position Angles (EVPAs) and significant $RM$ transverse gradient that provide strong evidence of a helical configuration of the magnetic field, as found in simulations.
}
    {}
\titlerunning{Multifrequency simultaneous VLBA view of the radio source 3C\,111}
\authorrunning{V. Bartolini et al.}
\maketitle

\section{Introduction} \label{sec:introduction}
Active Galactic Nuclei (AGN) are some of the most interesting astronomical objects among those observable using the Very Long Baseline Interferometer (VLBI) technique. 
AGN are excellent laboratories for a variety of physical processes. The standard model of AGN suggests the presence of a supermassive black hole (SMBH) at the center of the galaxy, which accretes matter and can be surrounded by an obscuring torus \citep{Urry_1995}. The complex dynamics around the SMBH and the presence of intense magnetic fields may lead to the production of two relativistic plasma streams of charged particles, emitting non-thermal radiation, called jets, propagating in opposite directions to each other.
Radio Galaxies (RGs) are defined as AGN with jets at a large angle to the line of sight and, among RGs, \citet{Faranoff_Riley} suggested a division based on their morphology and radio luminosity FR I radio galaxies have lower power (i.e. $P_\mathrm{1.4 \ \un{GHz}} < 10^{24.5} \ \un{W/Hz}$) and their emission on kpc scales is dominated by their jet; on the other hand, FR II radio galaxies, have high power (i.e. $P_\mathrm{1.4 \ \un{GHz}} > 10^{24.5} \ \un{W/Hz}$) and are dominated on large scales by hotspots, i.e. regions in which a jet terminates due to an interaction with the intergalactic medium (IGM).
The emission from RGs spans across the whole electromagnetic spectrum, from the radio domain to $\gamma$-rays \citep[e.g.][]{Casadio_2015,Grandi_2012}. 
Non-thermal radiation is mainly produced by charged particles, i.e., leptons interacting with each other (e.g., via scattering processes) and with the magnetic field in the jet (e.g., synchrotron emission). Therefore, properly characterizing the particle population and the magnetic field intensity and distribution is crucial to understanding the physics of AGN.
Thanks to the VLBI technique, it is possible to resolve radio jets down to event horizon scales \citep{EHT_2019}, and, therefore, it is one of the best tools to study the jet structure and evolution. \\
Synchrotron emission from AGN jets is polarized. Therefore, with a simultaneous multifrequency dataset, it is possible to infer the Rotation Measure structure \citep[see e.g.][for a complete review of the topic]{Hovatta_2012} and thus the Faraday-corrected magnetic field distribution in a certain region of the jet. \\
Relativistic magnetohydrodynamic (RMHD) simulation of AGN jets can successfully reproduce different magnetic field configurations based on different types of environmental and intrinsic AGN properties \citep[e.g.][]{Perucho_2019}. Hence, high-resolution VLBI observations of polarized AGN jets, along with RMHD simulations, are fundamental to shed light on the accretion and ejection mechanisms from SMBHs.
In this paper, we focus on an interesting case study: the Broad Line Radio Galaxy (BLRG) 3C\,111. \\
In the radio domain, 3C\,111 is classified as an FR II radio galaxy presenting two radio lobes and a single-sided jet. The jet lies at an angle of $\approx 17 \degree$ with respect to the line of sight \citep{Jorstad_2017}. On the parsec-scales, the counter-jet is not detected, likely due to Doppler deboosting \citep{Chatterjee_2011}. The radio core is highly compact and bright, and on the parsec-scales, the jet has a blazar-like behaviour, showing apparent superluminal motion \citep[e.g.][]{Beuchert_2018}, at odds with the kpc-scale morphology \citep{Linfield_1984}.
In addition to these features, given its proximity, VLBI observations of 3C\,111 $(z=0.049)$ \citep{Veron_2006} down to mm wavelengths enable us to probe the jet base with high spatial resolution and reduced opacity effects, and thus, to investigate the magnetic field geometry and nuclear environment within the jet collimation region \citep{Kovalev_2020}.
\\
The main aim of this work is to study the physics and magnetic field structure in the pc-scale jet of 3C\,111. This source has been widely studied throughout the years, both from an observational point of view \citep[e.g.][]{Kadler_2008, Beuchert_2018,Schulz_2020} and with the help of simulations \citep[e.g.][]{Perucho_2008,Fichet_2022}. Our work fits into this context thanks to our simultaneous observation that span across six different frequencies, providing the chance to study the jet structure from pc to sub-pc scales. 
This paper is organized as follows: 
In Sec. \ref{sec:Data_set}, we describe the calibration of the dataset used in this work; In Sec. \ref{sec:Results}, we produce maps for all the available frequencies, both in total and polarization intensity. After correcting for the core-shift, we produce the spectral index distribution for all the frequency pairs; In Sec. \ref{sec:Discussion} we explore the relation between the physical parameters associated with the different \texttt{modelfit} components (e.g. the magnetic field) and we retrieve Rotation Measure (RM) maps for two different triplets of frequency. \\
In this paper, we assume a $\Lambda$CDM cosmology with $H_\mathrm{0} = 69.6 \ \un{km \ s^{-1} \ Mpc^{-1}}$, $\Omega_\mathrm{M} = 0.286$, $\Omega_\mathrm{\Lambda} = 0.714$ \citep{Bennett_2014}.
At the redshift of 3C\,111, the luminosity distance is $D_\mathrm{L} = 214 \, \mathrm{Mpc}$, and the angular size of 1 mas corresponds to a linear size of $\approx 0.95$ pc.

\section{Data set and calibration} \label{sec:Data_set}
In this work, we analyze a set of multi-wavelength radio observations obtained with the Very Long Baseline Array (VLBA) on 08/05/2014. 
The dataset spans across 7 different frequencies: 1.4 GHz (L band), 5 GHz (C band), 8.4 GHz (X band), 15.2 GHz (U band), 21.9 GHz (K band), 43.8 GHz (Q band), and 87.6 GHz (W band) Each observing frequency is divided into 8 IFs with a bandwidth of $32 \un{MHz}$ and a channel separation of $500 \un{kHz}$, thus producing a total bandwidth per polarization of $256 \un{MHz}$. The total on-source time is $15.6 \un{m}$ for the C band, $9.1 \un{m}$ for the X band, $20.5 \un{m}$ for the U band, $147.9 \un{m}$ for the K band, $54.6 \un{m}$ for the Q band and $55.1$ for the W band. The recorded data rate at each band is $2 \un{Gbit/s}$.
Due to a reduced number of scans, we do not include the 1.4 GHz data in the analysis.
During the observations, North Liberty was not observing due to technical problems, leaving 9 telescopes observing.
To analyze and calibrate the data, we use \texttt{ParselTongue} \citep{ParselTongue}, a Python interface of AIPS, the Astronomical Image Processing System \citep{AIPS}. For the calibration of the dataset, we made use of a routine developed by J. L. Gómez and A. Fuentes that follows the standard procedures for polarization observations \citep[e.g.][]{Jorstad_2005}.
At this stage, it is possible to perform the imaging process with \texttt{Difmap} \citep{Difmap}. We used the CLEAN deconvolution algorithm \citep{CLEAN} along with self-calibration procedures in order to produce the total intensity images. 
After producing the spectrum of 3C\,111 integrated over the whole source, we noticed that the $21.9 \ \un{GHz}$ flux was lower than expected.
We inspected the integrated spectrum for other known flux density calibrators (e.g., 3C\,84) and found the same scaling issue. Namely, the K-band flux density was $\approx 20 \%$ lower than expected in all the sources, thus suggesting a possible issue in the observation. We noticed that for some scans, some baselines (e.g., the ones involving Brewster or Hancock) were dragging the average amplitude down. Therefore, we flagged those baselines and proceeded with the amplitude self-calibration without said antennas. Then, we used the obtained model to scale all the antennas.
We then performed the polarimetry calibration by uploading the total intensity images in \texttt{ParselTongue}. 
We use the AIPS task \texttt{LPCAL} to determine the instrumental polarization generally referred to as \textit{D-terms} polarization leakages \citep[see][for an extended description]{Casadio_2019}. 
We find leakages in the range of a few \%, with values progressively higher with frequency, but all below $10 \%$. To correct for the absolute orientation of the Electric Vector Position Angles (EVPAs), we considered the appropriate frequency data from the FGAMMA program \citep[$5 \un{GHz}$ and $8.4 \un{GHz}$]{FGAMMA}, the MOJAVE database ($15.2 \un{GHz} $), and the VLBA-BU-BLAZAR Program \footnote{\url{http://www.bu.edu/blazars/BEAM-ME.html}} for the $43 \un{GHz} $. We interpolated the data for the EVPA at $21.9 \un{GHz} $ and extrapolated those at $ 87.6  \un{GHz}$, the latter being, therefore, quite uncertain and to be handled with care. \\ 
In order to perform a discrete analysis of the jet emission, we modeled its brightness distribution at each frequency as a set of 2D circular Gaussian components through the \texttt{modelfit} function in \texttt{Difmap} following the approach by \citet{Lico_2012}.
This was done to better characterize the jet structure and to try to recognize the same features at different frequencies (see Sec. \ref{sec:comprel}).
To prevent the fitting procedure from modeling components that are not physically realistic in order to reach a better convergence, we fix the size of each component to $0.1$ of the beam minor axis whenever a component is shrunk to a size smaller than $10 \%$ of the beam minor axis. When the latter happens, we treat the size of such components as upper limits.
We estimate the uncertainties of the \texttt{modelfit} flux $\Delta F(\nu)$ with the Eq. \eqref{eq:flux_err}:

\begin{equation}
\label{eq:flux_err}
    \Delta F(\nu) = \sqrt{(0.1 \ F(\nu))^2 + (3\sigma_\mathrm{rms})^2}
\end{equation}
where the first term $(10\% \ F(\nu))$ is a conservative choice that represents the calibration error and $3\sigma_\mathrm{rms}$ is 3 times the off-source rms of the map, representing the statistical error as done in previous studies \citep[e.g.][]{Lico_2012}. 
The uncertainties on the position and size of the components are estimated based on the approach followed by \citet{Orienti_2011}. Namely, for the position, we estimate $\Delta r$ as $\Delta r = \theta / \mathrm{SNR}$, where $\theta$ is the angular size of the component, $\mathrm{SNR} = F(\nu) / (\sigma_\mathrm{rms} \times (\theta / \theta_{\mathrm{beam,min}}))$ is the signal-to-noise ratio and $\theta_{\mathrm{beam,min}}$ is the beam minor axis.
If $SNR > 10$, then the error is taken as $(1/10)\times \theta_{\mathrm{beam,min}}$.
We conservatively estimate the error of each component's size as $\Delta \theta = 0.1 \times \theta$.
The proper estimation of the errors for the \texttt{modelfit} components is non-trivial and, therefore, our work treats said uncertainties conservatively in order to ensure more robustness and reliability in our results. Comparing these with other work in the literature \citep[e.g.][]{Jorstad_2017,Lister_2009, Homan_2002}, our estimates are $\approx 1-2$ times larger.
\section{Results} \label{sec:Results}

\subsection{Total Intensity Images} \label{sec:Total}
\begin{figure*}
\begin{subfigure}{0.45\textwidth}
    \includegraphics[width=\textwidth]{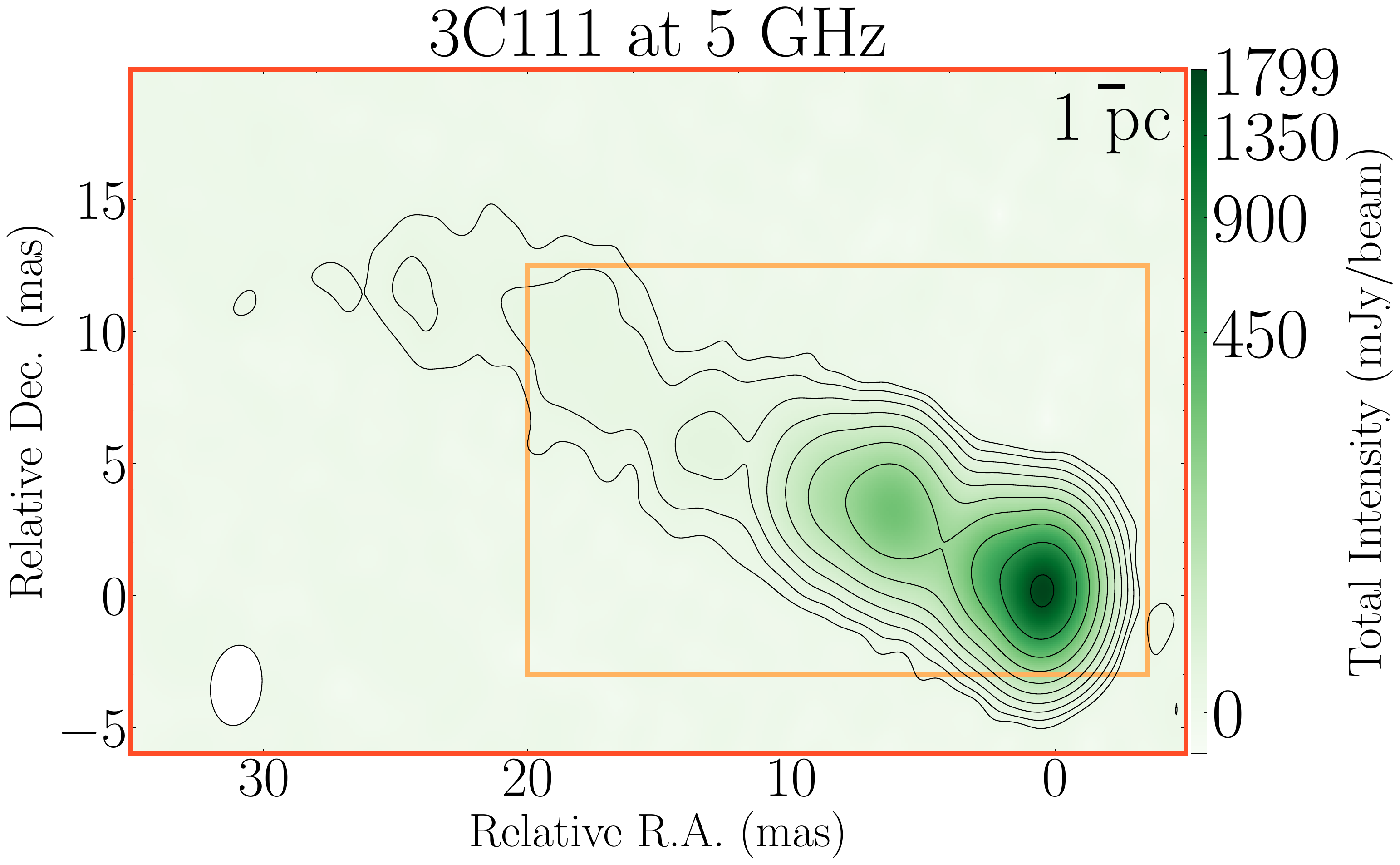}
    \caption{}
    \label{fig:totintc}    
\end{subfigure}
\hfill
\begin{subfigure}{0.45\textwidth}
    \includegraphics[width=\textwidth]{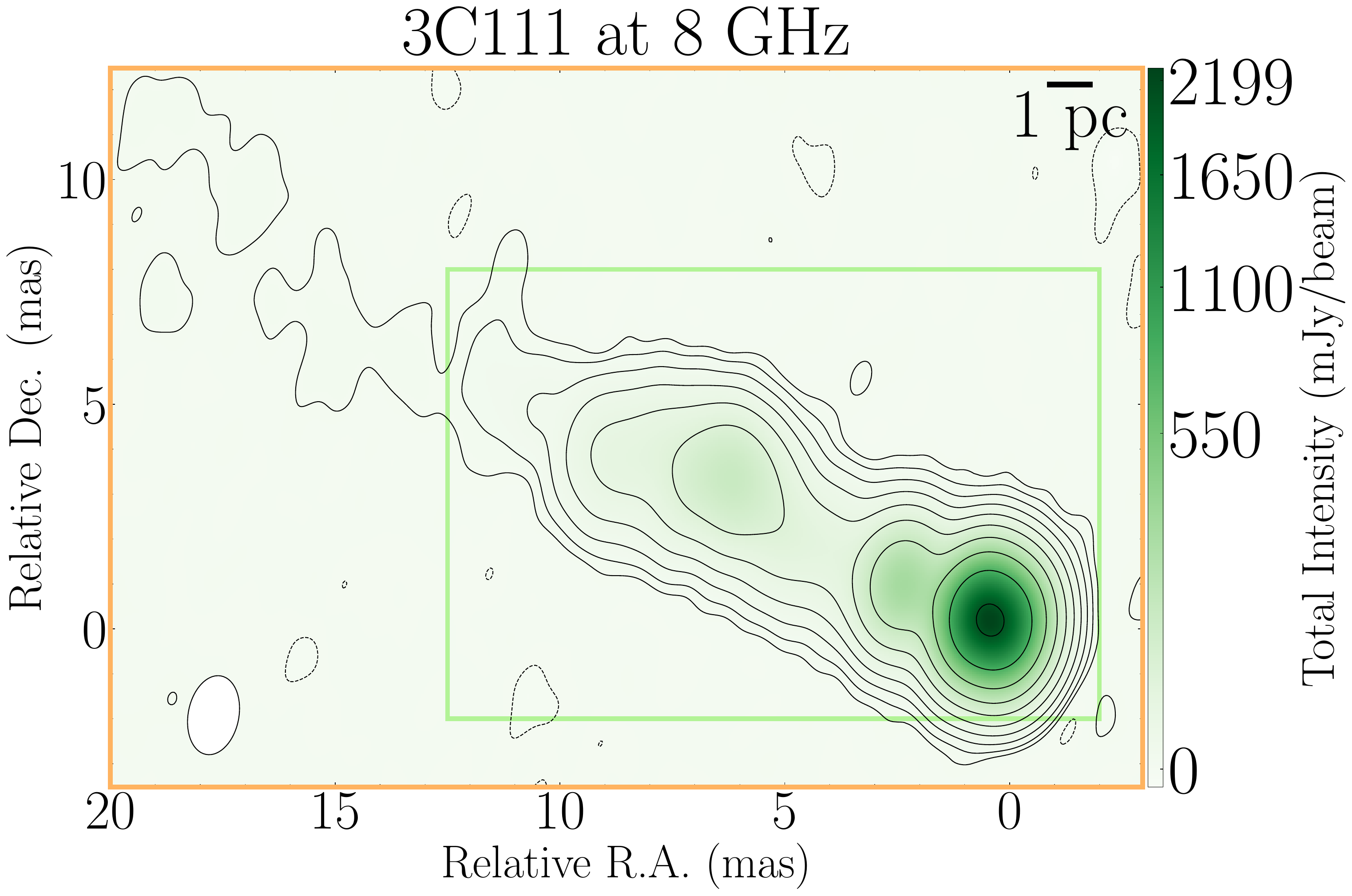}
    \caption{}
    \label{fig:totintx}    
\end{subfigure}
\hfill
\begin{subfigure}{0.45\textwidth}
    \includegraphics[width=\textwidth]{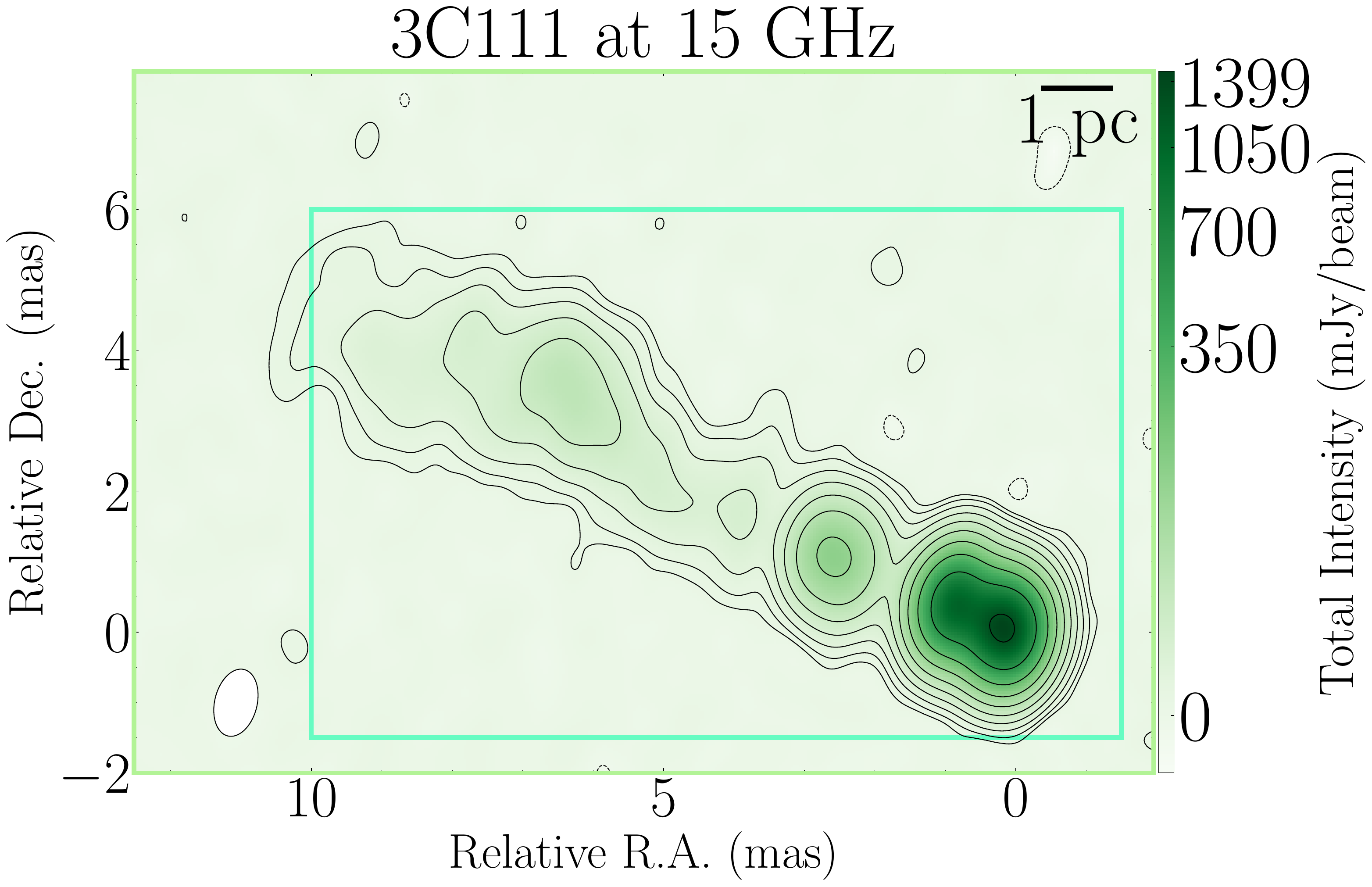}
    \caption{}
    \label{fig:totintu}    
\end{subfigure}
\hfill
\begin{subfigure}{0.45\textwidth}
    \includegraphics[width=\textwidth]{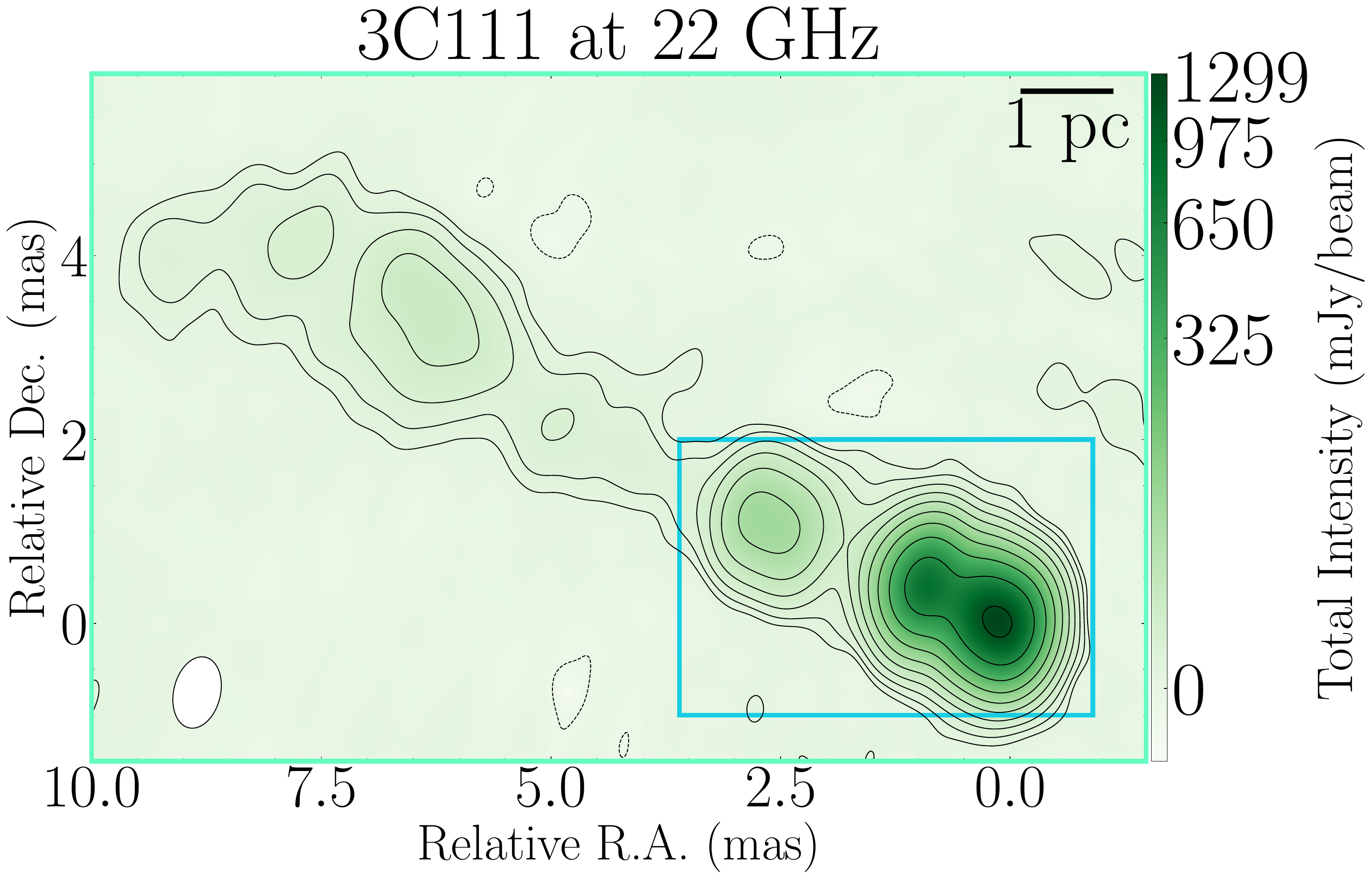}
    \caption{}
    \label{fig:totintk}    
\end{subfigure}
\hfill
\begin{subfigure}{0.45\textwidth}
    \includegraphics[width=\textwidth]{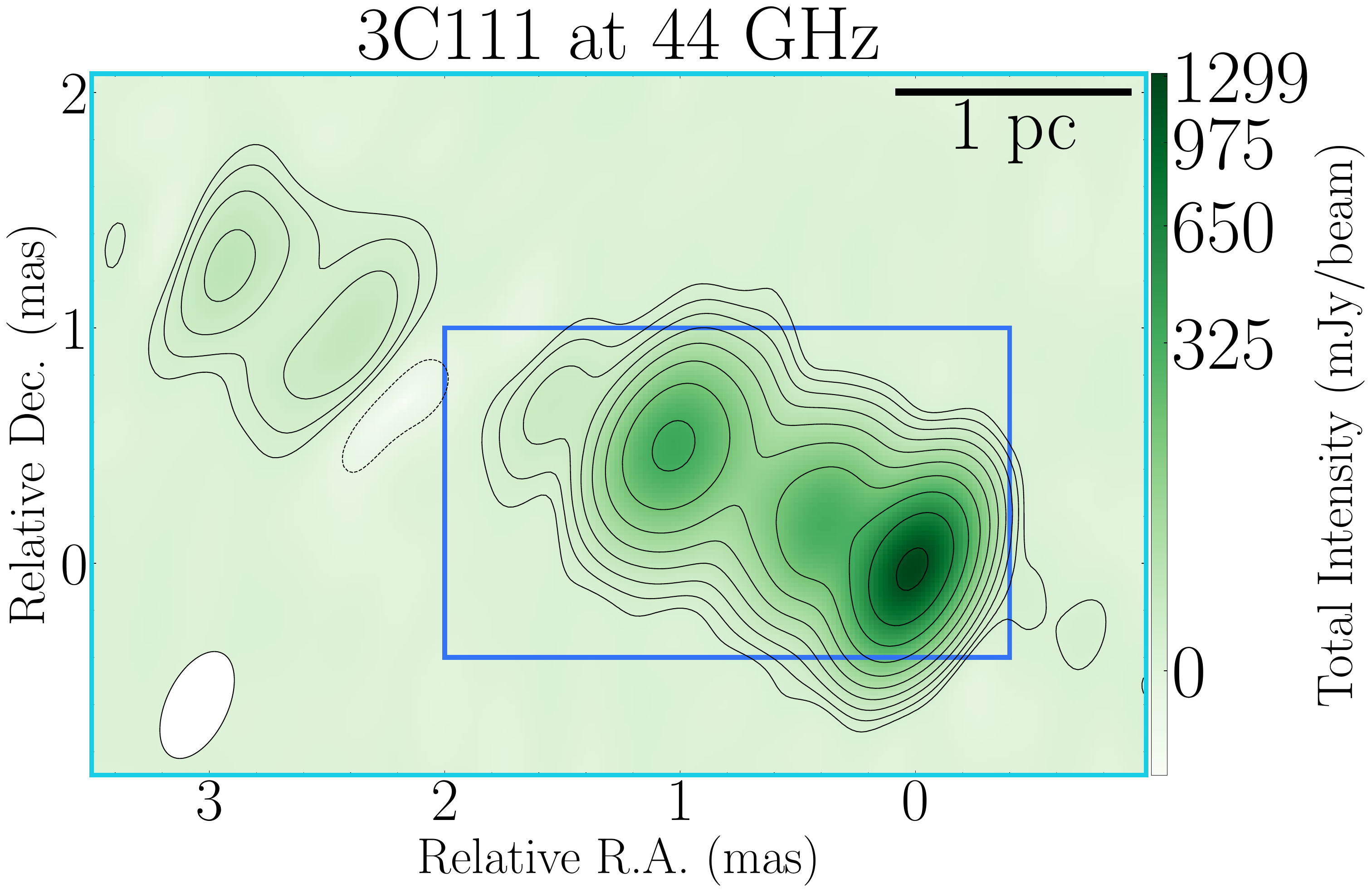}
    \caption{}
    \label{fig:totintq}    
\end{subfigure}
\hfill
\begin{subfigure}{0.45\textwidth}
    \includegraphics[width=\textwidth]{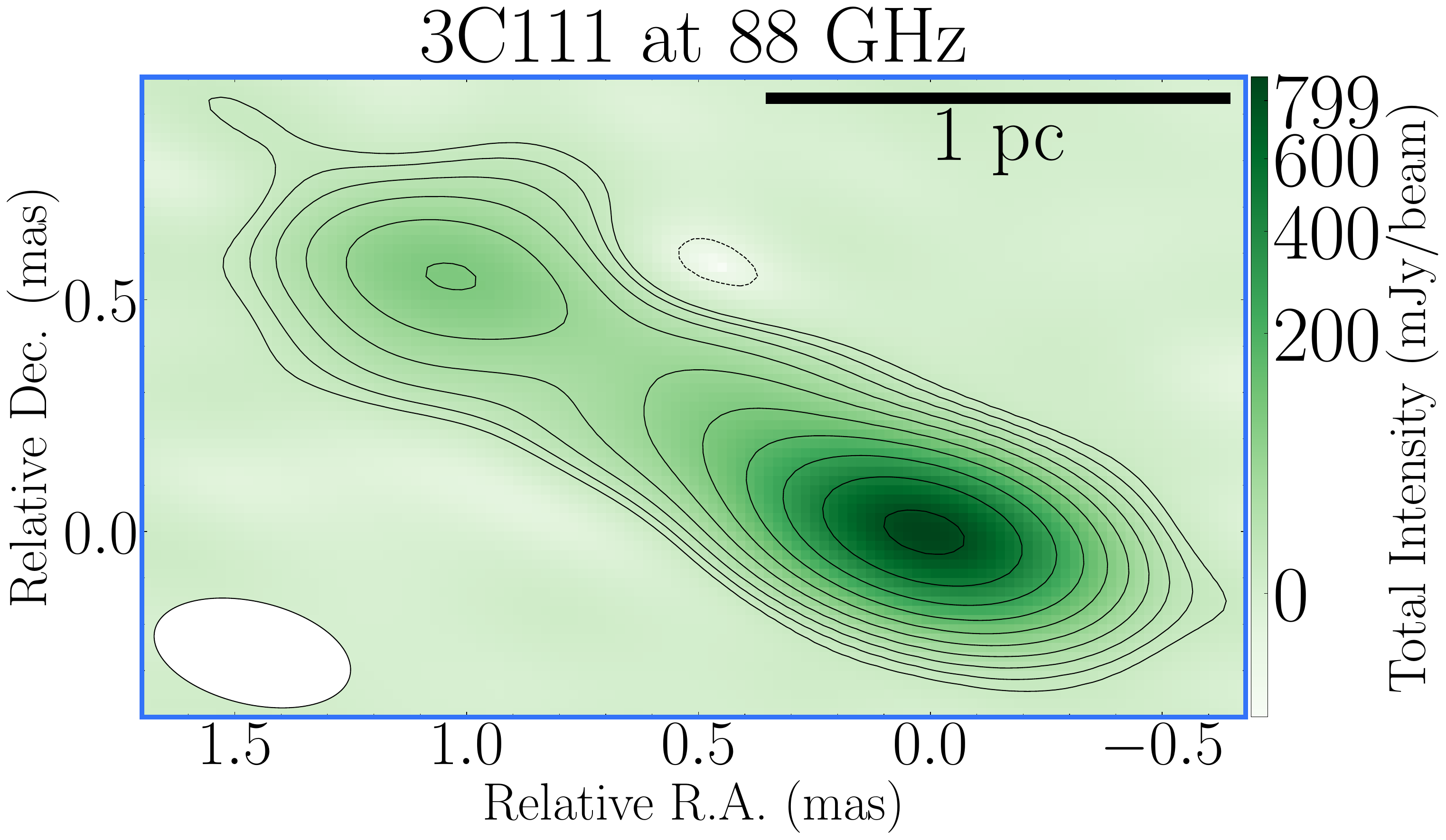}
    \caption{}
    \label{fig:totintw}    
\end{subfigure}
\hfill
\caption{Total intensity images of 3C\,111 for the 08/05/2014 observation produced in natural weighting. In each panel is plotted the restoring beam as a white ellipse in the bottom left corner and a size scale reference of 1 pc in the top right corner. Starting from \ref{fig:totintc}, in each image, there is a colored box representing the size of the following frequency image. The frequencies are color-coded from the lowest $(5 \un{GHz})$ to the highest $(87.6  \un{GHz})$, going from red to blue. \newline
\small (a) 3C\,111 total intensity image at $5 \un{GHz}$. The beam size is $3.06 \un{mas} \times 1.93 \un{mas}$ at $-7 \degree$, the pixel size is $0.08 \un{mas}$, the peak is $1.80 \un{Jy/beam}$, the total intensity is $2.94 \un{Jy}$, the rms is $0.21 \un{mJy/beam}$, contours are at $-0.03,0.03,0.07,0.18,0.43, $ $1.05,2.56,6.23,15.18,36.96,90.0 \%$ of the peak. (b) 3C\,111 total intensity image at $8.4 \un{GHz}$. The beam size is $1.78 \un{mas} \times 1.13 \un{mas}$ at $-11 \degree$, the pixel size is $0.04 \un{mas}$, the peak is $2.28 \un{Jy/beam}$, the total intensity is $3.84 \un{Jy}$, the rms is $0.19 \un{mJy/beam}$, contours are at $-0.04,0.04,0.09,0.22,0.52,$ $1.23,2.9,6.84,16.15,38.12,90.0 \%$ of the peak. (c) 3C\,111 total intensity image at $15.2 \un{GHz}$. The beam size is $0.97 \un{mas} \times 0.60 \un{mas}$ at $-14 \degree$, the pixel size is $0.04 \un{mas}$, the peak is $1.46 \un{Jy/beam}$, the total intensity is $3.49 \un{Jy}$, the rms is $0.23 \un{mJy/beam}$, contours are at $-0.06,0.06,0.14,0.31,0.7, $ $1.57,3.53,7.94,17.83,40.06,90.0 \%$ of the peak. (d) 3C\,111 total intensity image at $21.9 \un{GHz}$. The beam size is $0.79 \un{mas} \times 0.51 \un{mas}$ at $-13 \degree$, the pixel size is $0.02 \un{mas}$, the peak is $1.31 \un{Jy/beam}$, the total intensity is $3.33 \un{Jy}$, the rms is $0.46 \un{mJy/beam}$, contours are at $-0.15,0.15,0.3,0.62,1.26,$ $2.57,5.23,10.65,21.69,44.18,90.0 \%$ of the peak. (e) 3C\,111 total intensity image at $43.8 \un{GHz}$. The beam size is $0.49 \un{mas} \times 0.26 \un{mas}$ at $-26 \degree$, the pixel size is $0.02 \un{mas}$, the peak is $1.31 \un{Jy/beam}$, the total intensity is $2.66 \un{Jy}$, the rms is $1.14 \un{mJy/beam}$, contours are at $-0.34,0.34,0.64,1.18,2.2,$ $4.08,7.57,14.06,26.1,48.47,90.0 \%$ of the peak. (f) 3C\,111 total intensity image at $ 87.6  \un{GHz}$. The beam size is $0.43 \un{mas} \times 0.22 \un{mas}$ at $77 \degree$, the pixel size is $0.02 \un{mas}$, the peak is $0.89 \un{Jy/beam}$, the total intensity is $1.18 \un{Jy}$, the rms is $2.13 \un{mJy/beam}$, contours are at $-0.84,0.84,1.42,2.38,4.0,$ $6.72,11.29,18.97,31.88,53.56,90.0 \%$ of the peak.}
\label{fig:totint}
\end{figure*}
We produce total intensity images, shown in Fig. \ref{fig:totint}, for all  6 independent frequencies available in the dataset. For each of them, we plot the restoring beam in the bottom left corner and also the length of $1$ pc in the top right corner in order to provide a visualization of the spatial resolution achieved. 
The source is well detected at all frequencies, dominated by a compact core of brightness ranging between $\approx 1 \ \un{Jy}$ and $\approx 3 \ \un{Jy}$. 
From the core, a one-sided jet emerges, extending up to $\approx 25 \ \un{mas}$ in length in the lowest frequency image (e.g., $5 \ \un{GHz}$).  The jet is remarkably straight, being oriented at a position angle of $\approx 65 \degree$, in agreement with previous studies of 3C\,111 \citep[e.g.][]{Kadler_2008,Beuchert_2018,Schulz_2020} and shows no evidence of any bending over its full extension. At the highest frequency of $87.6  \un{GHz}$, we achieve a resolution of $0.22 \un{mas}$.

\subsection{Core-shift analysis}\label{sec:core_shift}
\begin{figure}
    \centering
    \includegraphics[width=\columnwidth]{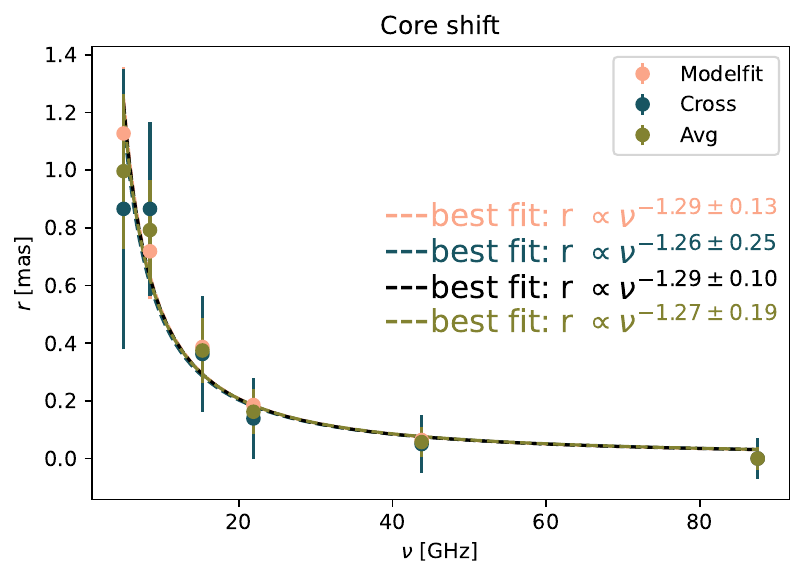}
    \caption{The core-shift effect for 3C\,111 between all the 6 frequencies taken into account in this work (from 5 GHz to 87.6 GHz). The different colors represent the different methods used to estimate the core shift, as explained in \ref{sec:core_shift}. The black line $r \propto \nu^{-1.29 \pm 0.10}$ represents a fit of all the points together. The fit performed with the AVG data points gives $r\propto \nu^{-1.27 \pm 0.19}$, implying $k_{\mathrm{r}} = 0.79 \pm 0.19$. This result is suggestive of a mild dominance of the particle over the magnetic field, in the energy budget. }
    \label{fig:coreshift}
\end{figure}

The radio core is the region at the base of a jet where there is a transition from an optically thick regime to an optically thin one. Therefore, the observed position of the radio core depends on the frequency observed
\citep[e.g.][]{Lobanov_1998}. 
The transition region (i.e. the core) changes with frequency as $r \propto \nu^{1/k_\mathrm{r}}$, where $\nu$ is the frequency and $k_\mathrm{r}$ is the power index that, in the condition of equipartition between jet particle and magnetic field energy densities, is equal to $-1$.
To study the core-shift, we produce and analyze the maps in two different ways. The first method (Fits) is based on the 2D cross-correlation script FITSAlign, described in \citet{MOJAVE_Push_2012}, and the second one, labeled as Modelfit, is based on the \texttt{modelfit} components. 
For the first method, we compute the Equivalent Beam following Eq. \eqref{eq:eq_beam} and produce the maps with a pixel size of $\frac{1}{7} EB$.
Even though \citet{Fromm_2013} claimed that the best choice for the pixel size is 1/20 of the high-frequency beam size, we opted for a higher value. This is done for both avoiding incurring into over-resolution issues and to have comparable error bars with the \texttt{Modelfit} method.
\begin{equation}\label{eq:eq_beam}
    EB = \sqrt{B_{min} \cdot B_{max}} 
\end{equation}
where $B_{min}$ and $B_{max}$ are the minor and major axes of the restoring beam.
We then produce the maps at two consecutive frequencies with the same imaging parameters (see Sec. \ref{sec:Spectral}). We choose the common pixel size as $\frac{1}{7} CB$, where $CB$ is the common beam, i.e., the mean between the two EBs.
The FITSAlign script is based on the 2D pixel-based cross-correlation of a user-selected optically thin region. The program calculates a normalized cross-correlation between the low-frequency image and the selected feature of the high-frequency image using a fast method described by \citet{Lewis_1995}.
For each $\nu$, we sum the obtained shift with all the ones at higher frequencies in order to reference the core-shift values to the $87.6  \ \un{GHz}$ core. 
As described in \citet{MOJAVE_Push_2012}, the core shift is the offset between two maps, summed with the difference between the relative peaks in each map. Thus, after shifting the maps, we add the relative difference between the peaks, computed with \texttt{Difmap}.
We compute the associated uncertainties for $r$, through the Eq. \eqref{eq:coreerr}:
\begin{equation}\label{eq:coreerr}
    \Delta r = \sqrt{\left(\dfrac{x}{\sqrt{x^2 + y^2}} \Delta x\right)^2 + \left(\dfrac{y}{\sqrt{x^2 + y^2}} \Delta y\right)^2} 
\end{equation}
where $x$ and $y$ are the x and y position  of the peak, in $\un{mas}$. 
\begin{table*}
    \caption{\label{tab:coreschift} Core-shift results for the two methods (Modelfit, Cross) described in Sec. \ref{sec:core_shift}, and their average value (AVG).}
    \resizebox{\textwidth}{!}{%
    \begin{tabular}{l!{\vrule width 0.5mm}c|c|c!{\vrule width 0.5mm}c|c|c!{\vrule width 0.5mm}c|c|c}
     & \multicolumn{3}{c!{\vrule width 0.5mm}}{Modelfit} & \multicolumn{3}{c!{\vrule width 0.5mm}}{Cross} & \multicolumn{3}{c}{AVG} \\
    \hline
    $\nu\un{(GHz)}$ & $r\un{(mas)}$ & $x\un{(mas)}$ & $y\un{(mas)}$ &  $r\un{(mas)}$ & $x\un{(mas)}$ & $y\un{(mas)}$ & $r\un{(mas)}$ & $x\un{(mas)}$ & $y\un{(mas)}$ \\
    \midrule
5.0 & $ 1.13 \pm 0.23 $ & $ 1.02 \pm 0.21 $ & $ 0.48 \pm 0.10 $ & $ 0.87 \pm 0.49 $ & $ 0.80 \pm 0.34 $ & $ 0.33 \pm 0.34 $ & $ 1.00 \pm 0.27 $ & $ 0.91 \pm 0.20 $ & $ 0.40 \pm 0.18 $ \\ 

8.4 & $ 0.72 \pm 0.17 $ & $ 0.64 \pm 0.15 $ & $ 0.32 \pm 0.08 $ & $ 0.87 \pm 0.30 $ & $ 0.80 \pm 0.21 $ & $ 0.33 \pm 0.21 $ & $ 0.79 \pm 0.17 $ & $ 0.72 \pm 0.13 $ & $ 0.33 \pm 0.11 $ \\ 

15.2 & $ 0.39 \pm 0.11 $ & $ 0.35 \pm 0.10 $ & $ 0.16 \pm 0.05 $ & $ 0.36 \pm 0.20 $ & $ 0.32 \pm 0.14 $ & $ 0.17 \pm 0.14 $ & $ 0.37 \pm 0.11 $ & $ 0.34 \pm 0.09 $ & $ 0.17 \pm 0.07 $ \\ 

21.9 & $ 0.19 \pm 0.07 $ & $ 0.17 \pm 0.07 $ & $ 0.08 \pm 0.03 $ & $ 0.14 \pm 0.14 $ & $ 0.12 \pm 0.10 $ & $ 0.07 \pm 0.10 $ & $ 0.16 \pm 0.08 $ & $ 0.14 \pm 0.06 $ & $ 0.07 \pm 0.05 $ \\ 

43.8 & $ 0.06 \pm 0.04 $ & $ 0.06 \pm 0.04 $ & $ 0.03 \pm 0.02 $ & $ 0.05 \pm 0.10 $ & $ 0.05 \pm 0.07 $ & $ 0.00 \pm 0.07 $ & $ 0.06 \pm 0.05 $ & $ 0.05 \pm 0.04 $ & $ 0.01 \pm 0.04 $ \\ 

87.6 & $ 0.00 \pm 0.03 $ & $ 0.00 \pm 0.03 $ & $ 0.00 \pm 0.02 $ & $ 0.00 \pm 0.07 $ & $ 0.00 \pm 0.05 $ & $ 0.00 \pm 0.05 $ & $ 0.00 \pm 0.04 $ & $ 0.00 \pm 0.03 $ & $ 0.00 \pm 0.03 $ \\

    \bottomrule
    \end{tabular}
    }
    \tablefoot{For each method, the relative columns represent the distance from the $87.6$ GHz core $r$ and the x and y shift from it ($x$ and $y$, respectively) in mas. The rows show the frequency taken into account in GHz.}    
    \label{tab:coreshift}

\end{table*}
Since $\Delta x = \Delta y = px$, the associated error on $r$ is again equal to the pixel size.
When summing the shifts, we propagate the uncertainties following the theory of error propagation with the help of the \texttt{uncertainties}\footnote{\url{http://pythonhosted.org/uncertainties/}} python package.
The second method is based on the \texttt{modelfit} components.
We shifted all the \texttt{modelfit} components in such a way that the core component at each frequency is located at the phase center of each map.
We then matched the same component at two consecutive frequencies following the procedure outlined in \ref{sec:comprel}.
We then computed the shift between the two frequencies for all the matched components. If two or more components were matched in the same frequency pair, we took the mean value for the shift. All the selected components lie in an extended region of the jet, which is optically thin. \\
Moreover, for every frequency, we computed the average (AVG) value between the two methods. The value of the parameters with their associated uncertainties is presented in Tab. \ref{tab:coreshift}.
The two methods and their average values yield similar results for the fitting values, as can be seen in Fig. \ref{fig:coreshift}.
The black line $r \propto \nu^{-1.29 \pm 0.10}$ represents a fit of all the points. All the values from different methods are consistent with each other. The fit in the AVG case gives $r \propto \nu^{-1.27 \pm 0.19}$, corresponding to $k_\mathrm{r} = 0.79 \pm 0.19$. This can be suggestive of a slight particle-dominance with respect to the magnetic field energy density.
We used the \texttt{scipy (v1.8.0)} function \href{https://docs.scipy.org/doc/scipy/reference/generated/scipy.optimize.curve_fit.html}{\texttt{curve\_fit}} to fit the parameters, taking into account the errors. 
The error of $k_\mathrm{r}$ has been estimated as the square root of the diagonal elements of the covariance matrix produced with the same \texttt{scipy} module.

\subsection{Spectral index}\label{sec:Spectral}
\begin{figure*}
    \centering
    \begin{subfigure}{0.45\textwidth}
        \includegraphics[width=\textwidth]{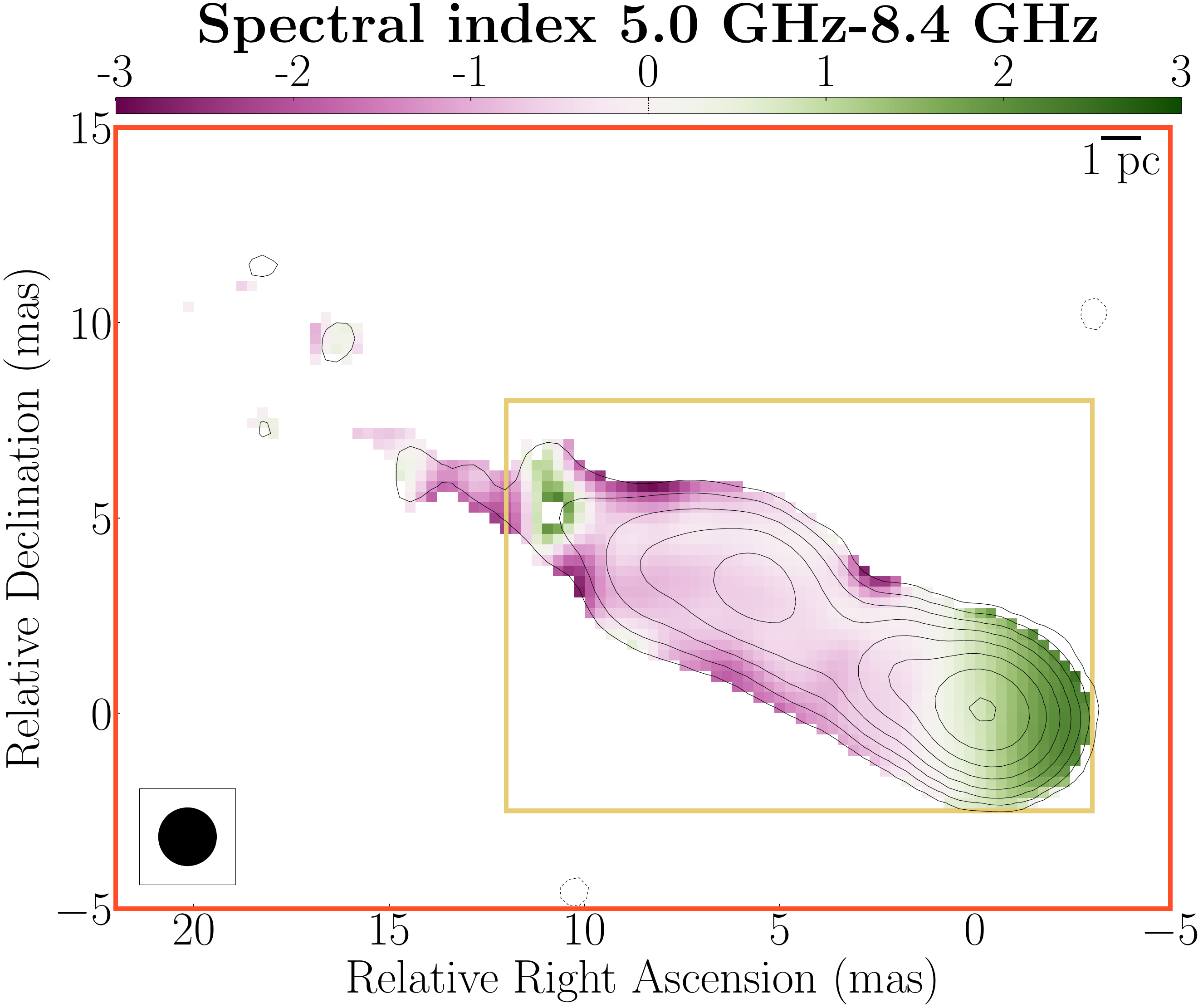}
        \caption{}
        \label{fig:spixcx}
    \end{subfigure}
    \hfill
    \begin{subfigure}{0.45\textwidth}
        \includegraphics[width=\textwidth]{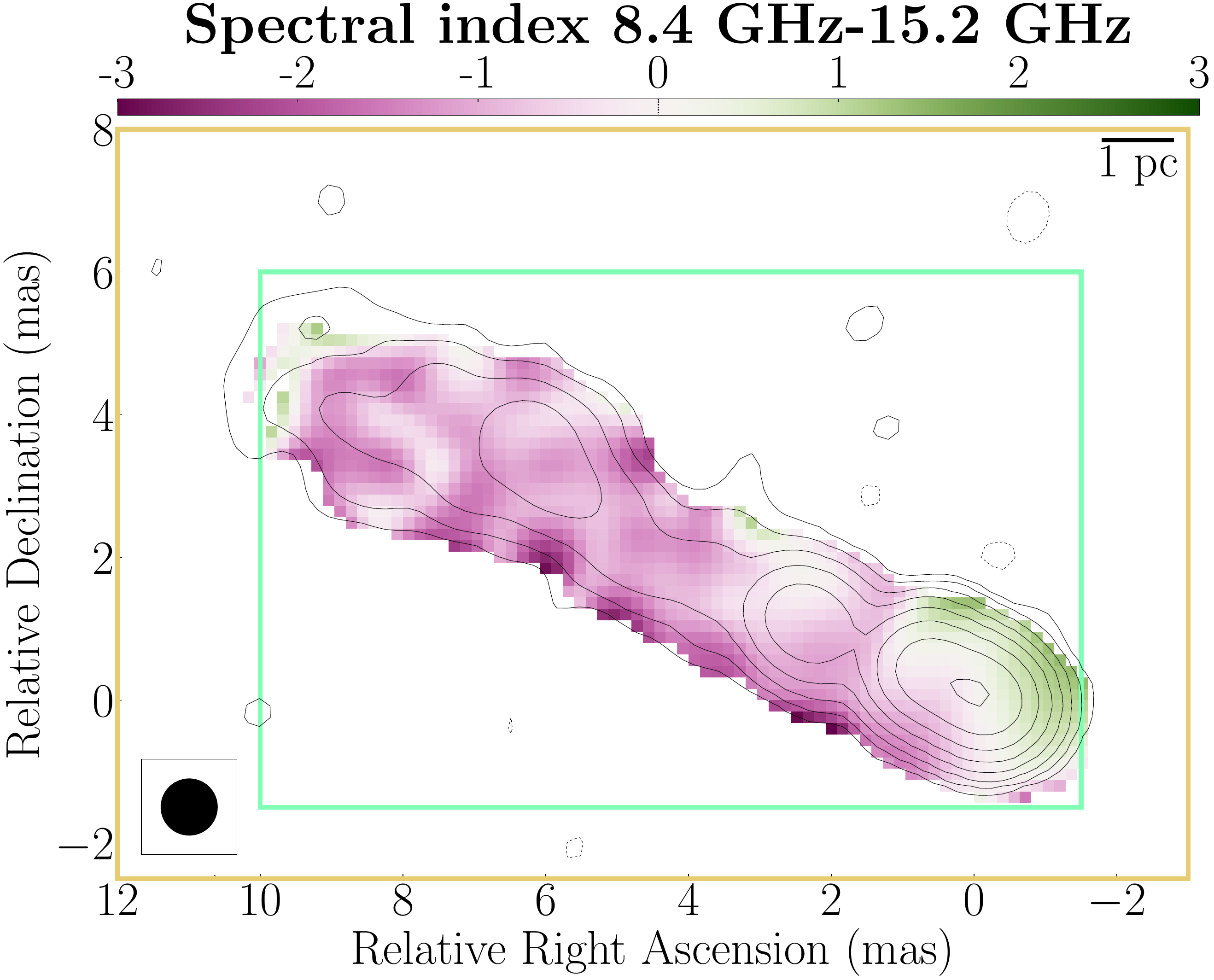}
        \caption{}
        \label{fig:spixxu}
    \end{subfigure}
    \hfill
    \begin{subfigure}{0.45\textwidth}
        \includegraphics[width=\textwidth]{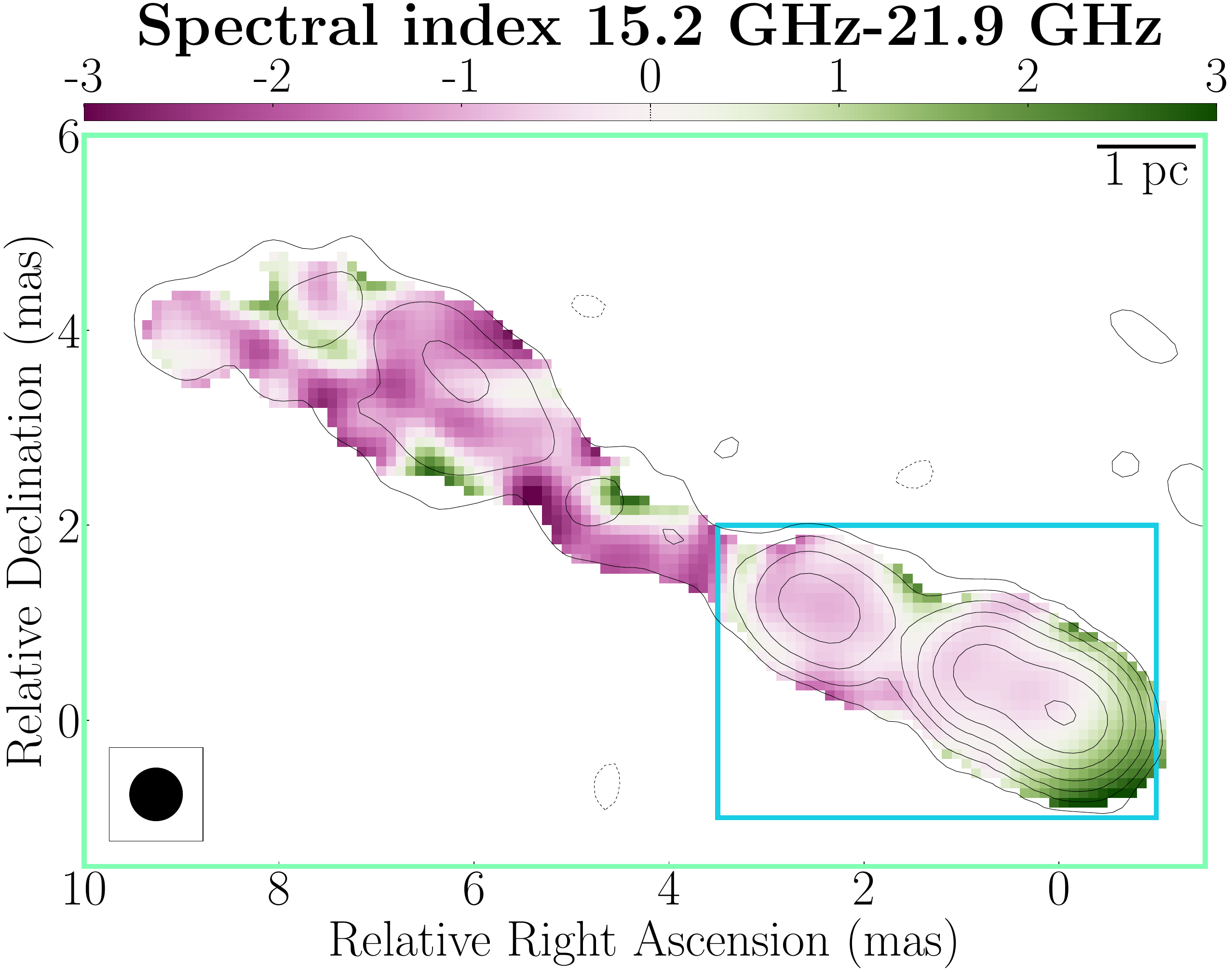}
        \caption{}
        \label{fig:spixuk}
    \end{subfigure}
    \hfill
    \begin{subfigure}{0.45\textwidth}
        \includegraphics[width=\textwidth]{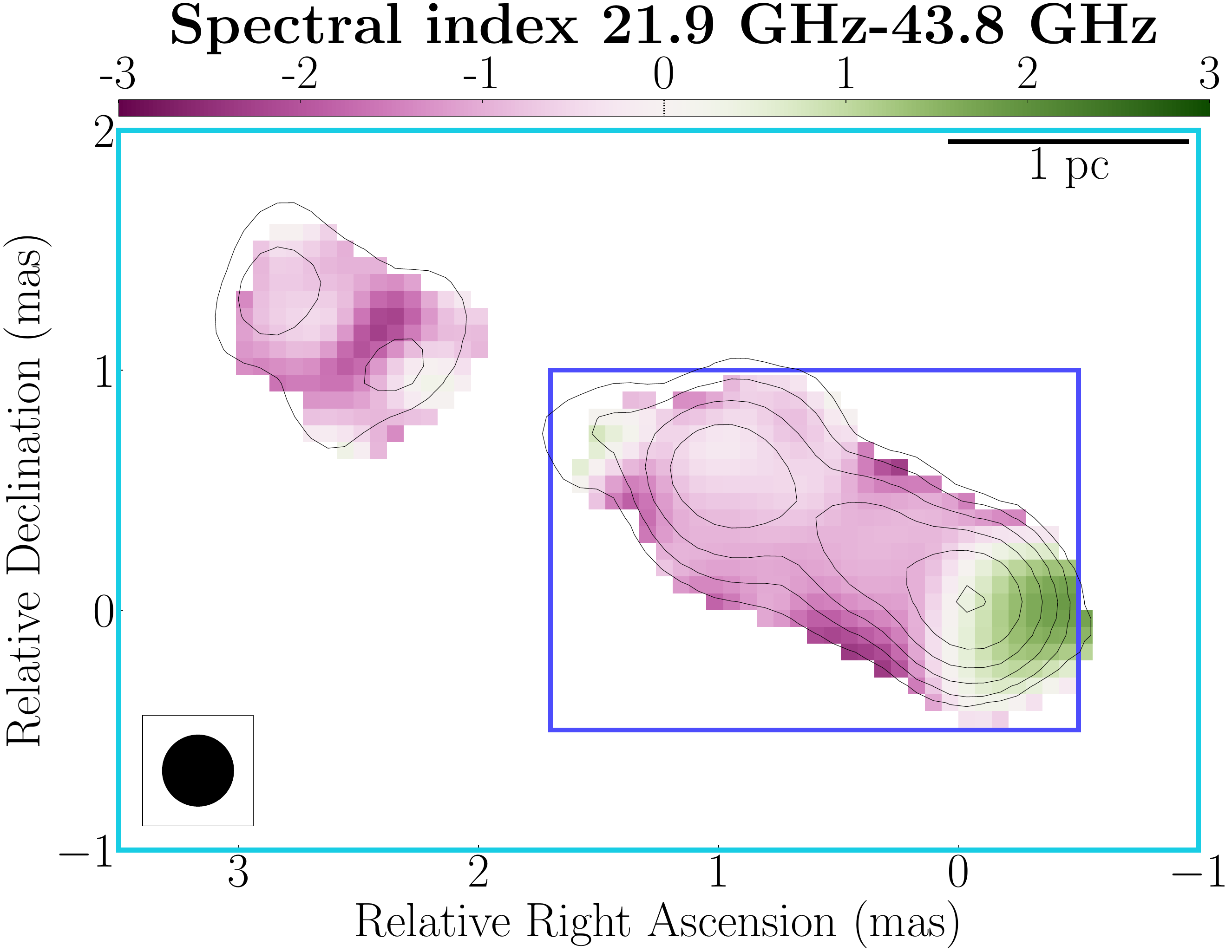}
        \caption{}
        \label{fig:spixkq}
    \end{subfigure}
    \hfill
    \begin{subfigure}{0.45\textwidth}
        \includegraphics[width=\textwidth]{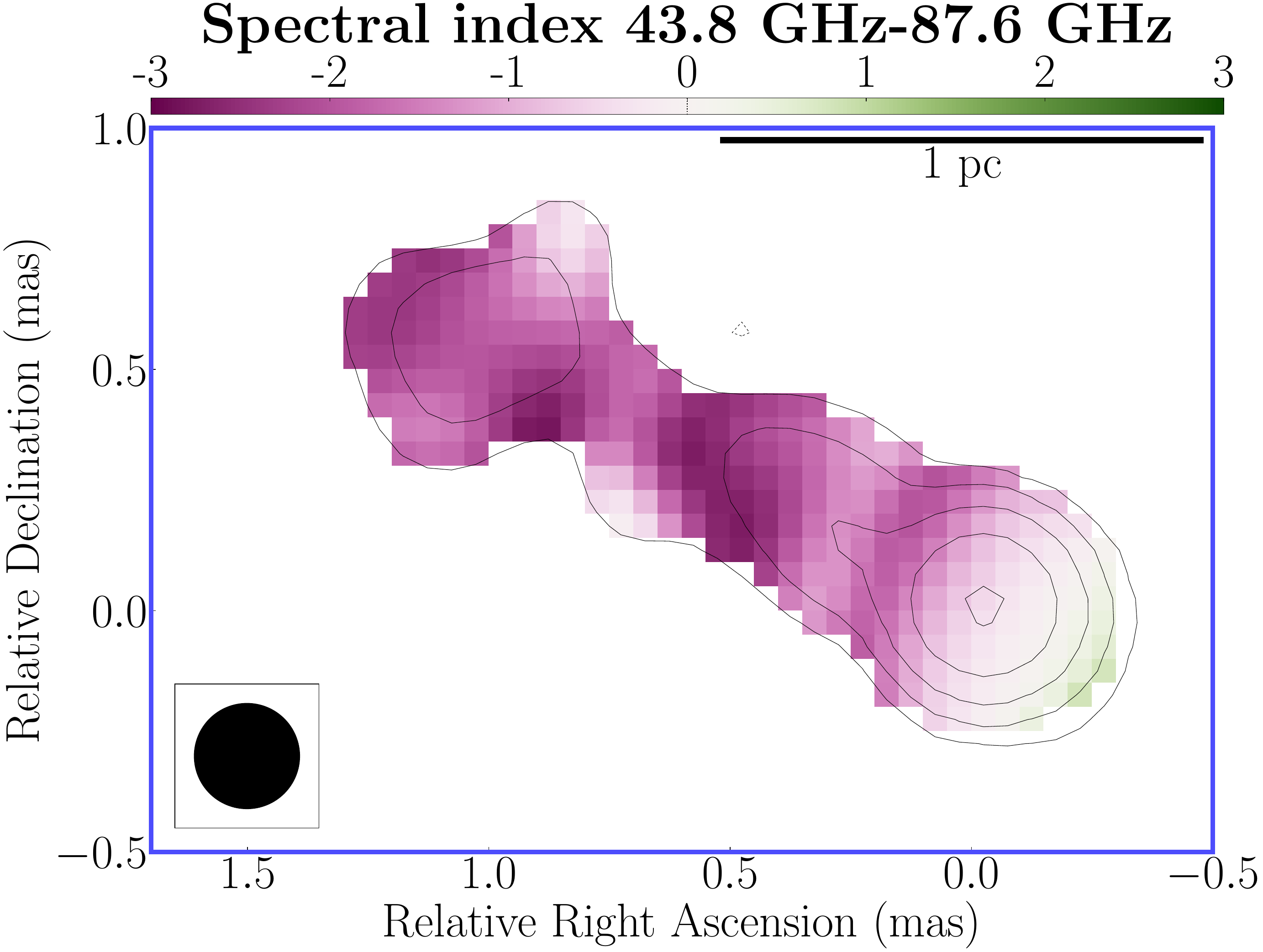}
        \caption{}
        \label{fig:spixqw}
    \end{subfigure}
    \hfill    
    \caption{Spectral index maps for all 5 frequency pairs, computed following Sec. \ref{sec:Spectral}, plotted over the higher frequency contours. In each panel, the common beam for the two frequencies is plotted as a black circle in the bottom left corner, and a size scale reference of 1 pc is in the top right corner. The frequency pairs are color-coded from the lowest $(5-8.4 \un{GHz})$ to the highest $(43.8-87.6  \un{GHz})$, going from red to blue. The color map is chosen from a colour-vision deficiency-friendly package \citep{Crameri_2021}. }.
    \label{fig:spix}
\end{figure*}

We produce spectral index ($S_\nu \propto \nu^{\alpha}$) maps for all 5 frequency pairs to investigate the distribution of $\alpha$ across the jet from parsec to sub-parsec scales. 
The images were produced for each frequency pair, imposing the same $uv_{min}$, $uv_{max}$, $mapsize$, $pixelsize$, and restoring $beam$. The relevant information is collected in Tab. \ref{tab:spix_param}
The spectral index maps are produced with the FITSAlign script (see Sec. \ref{sec:core_shift}) using a threshold of $5\times \sigma_{h}$, where $\sigma_{h}$ is the off-source rms of the higher frequency map.
\begin{table}
    \caption{\label{tab:spix_param} The chosen parameters of the spectral index maps are shown in  Fig. \protect\ref{fig:spix}.}
    \label{tab:spix_param}
    \centering
    \begin{tabular}{c c c c c }
    \hline\hline
    $\nu $ & UV & Beam  & Map & Pixel \\
    \hline 
        5-8.4 & $4.06$ - $146.25$ & $1.9 \times 1.9$ & $2048$ & $0.08$ \\ 
        8.4-15.2 & $7.54$ - $245$ & $1.1 \times 1.1$ & $2048$ & $0.04$ \\ 
        15.2-21.9 & $11.14$ - $441.5$ & $0.6 \times 0.6$ & $2048$ & $0.04$ \\ 
        21.9-43.8 & $23.5$ - $632$ & $0.5 \times 0.5$ & $2048$ & $0.02$ \\ 
        43.8-87.6  & $128.7$ - $686$ & $0.2 \times 0.2$ & $1024$ & $0.02$ \\ 
    \hline
    \end{tabular}
    \tablefoot{$\nu$ is the frequency pair in GHz; UV is the UV range in $\un{M\lambda}$; Beam is the size of the common beam in mas; Map is the size of the map in pixels and pixel is pixel size in mas.}
\end{table}
We use Eq. \ref{eq:spixerr} to estimate the uncertainties of the spectral index:
\begin{equation}
\label{eq:spixerr}
    \alpha_\mathrm{err} = \dfrac{1}{\log\dfrac{\nu_2}{\nu_1}}\sqrt{\left(\dfrac{\Delta F(\nu_1)}{F(\nu_1)}\right)^2 + \left(\dfrac{\Delta F(\nu_2)}{F(\nu_2)}\right)^2}
\end{equation}
where $\Delta F(\nu)$ represents the error of the flux density that has been estimated following Eq. \ref{eq:flux_err}. The calibration contribution dominates the total error of $\Delta F(\nu)$ as long as $F(\nu) \gg \sigma_\mathrm{rms}$. This holds throughout the jet extension, whereas, on the edges of the jet, the rms contribution starts to rise.
In the final spectral index images, all pixels with $SNR < 5$ were blanked.
All the spectral index maps are presented in Fig. \ref{fig:spix}.

\subsection{Model-fitting of the brightness distribution} \label{sec:Modelfit}
\begin{figure*}
    \centering
    \includegraphics[width=17cm]{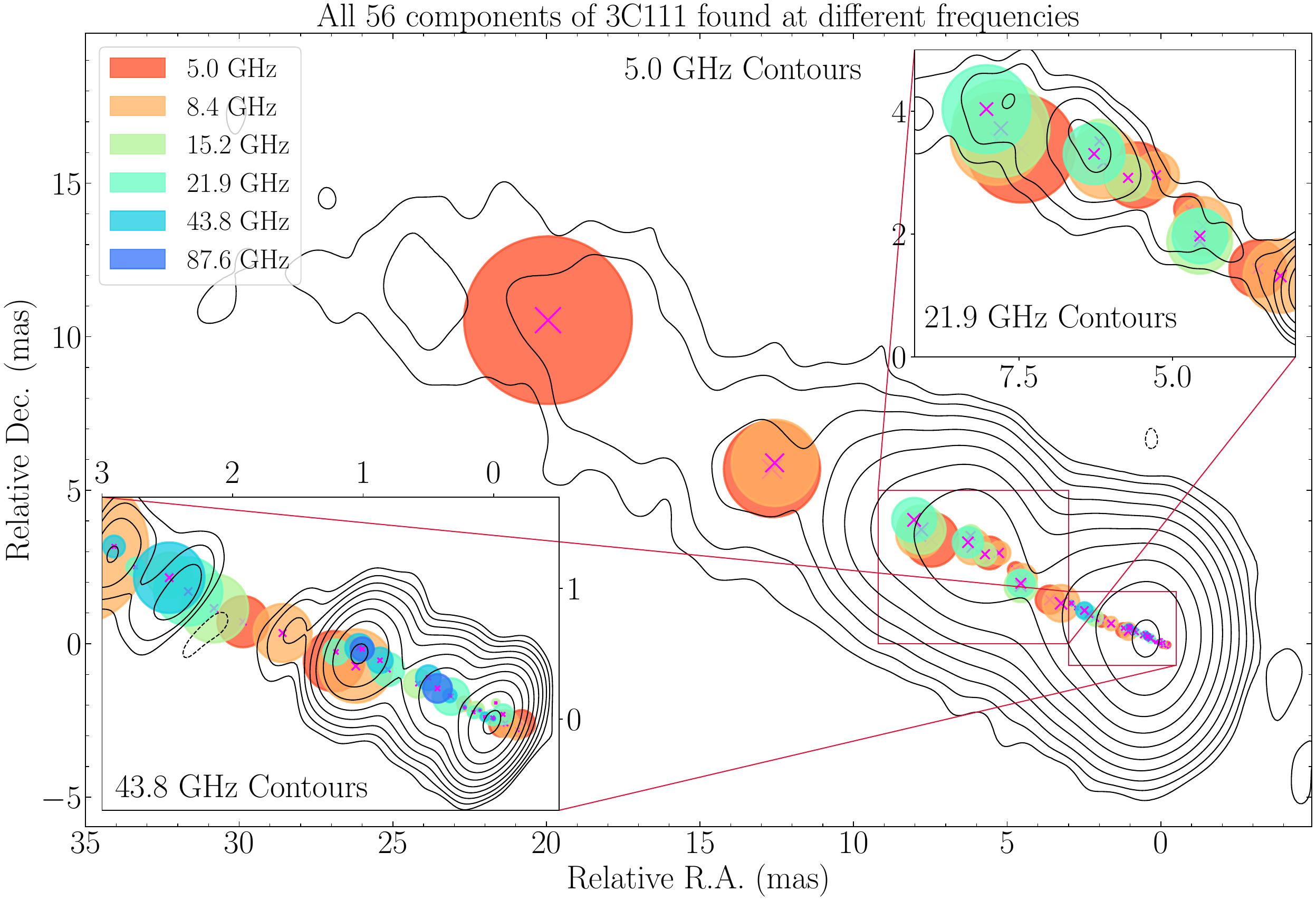}
    \caption{All the \texttt{modelfit} components found in 3C\,111, plotted over the 5 GHz contours (\ref{fig:totintc}). The components at each frequency are represented by the color in the legend, and their position is corrected for the core-shift found in Sec. \ref{sec:core_shift}.
    The upper right panel shows the region between $9$ and $2$ mas in Right Ascension and $0$ and $5$ in Declination. The contours are of the 21.9 GHz maps (\ref{fig:totintk}). The lower-left panel shows the region between $-0.5$ and $3$ mas in Right Ascension and $-0.7$ and $1.7$ in Declination. The contours are of the 43.8 GHz map (\ref{fig:totintq}).}
    \label{fig:allcomp}
\end{figure*}
As explained in Sec. \ref{sec:Data_set}, we used the \texttt{modelfit} function in \texttt{Difmap} to fit the visibilities with discrete 2D circular Gaussians. The total number of components found is 56, confirming the richness in the jet structure of 3C\,111.
\texttt{Modelfit} provides several parameters for each component: the flux density, the distance and position angle with respect to the center of the map, and the FWHM size of the Gaussian.
We could then compute the observed brightness temperature $T_{\mathrm{B}}$ and the equipartition magnetic field $H_{\mathrm{eq}}$ for each component.
The values found for all the components with the associated uncertainties are reported in Tab. \ref{tab:components}. 
We estimate the observed brightness temperature as:
\begin{equation}
\label{eq:T_comp}
    T_{\mathrm{B}} \approx 1.22 \times 10^{12} \dfrac{F(\nu)}{\un{Jy}} \left(\dfrac{\un{mas}}{\theta}\right)^2 \left(\dfrac{\un{GHz}}{\nu}\right)^2 (1+z) \ \un{K}
\end{equation}
where $F(\nu)$ is the flux density of the component in $\un{Jy}$, $\theta$ is its angular size in $\un{mas}$, $\nu$ is the observing frequency in $\un{GHz}$ and $z$ is the redshift $(=0.049)$ \citep{Veron_2006}.
We estimate the uncertainties for the brightness temperature with Eq. \eqref{eq:Tunc}:
\begin{equation}
\label{eq:Tunc}
    \Delta T_{\mathrm{B}} = T_{\mathrm{B}} \sqrt{\left(\dfrac{\Delta F(\nu)}{F(\nu)}\right)^2 + \left(\dfrac{2\Delta \theta}{\theta}\right)^2}
\end{equation}
For the core component, we find observed brightness temperature values in the range $0.1-1.3 \times 10^{12} \ \un{K}$ for the various frequencies. The different range of values can be ascribed to the fact that for various resolutions, the core region can be fitted with one or more \texttt{modelfit} components. 
We estimate the magnetic field under the assumption that it is at the minimum energy state (i.e., equipartition magnetic field) only for the matched components at two different frequencies (see \ref{sec:comprel}). Therefore, it is possible to compute the magnetic flux density through Eq. \eqref{eq:Heqeq}, taken from \citet{Pacholczyk_1970}:

\begin{equation}
\label{eq:Heqeq}
    H_{\mathrm{eq}} = \left(4.5 \dfrac{1+k}{\Phi} c_{12}\dfrac{L}{V}\right)^{2/7} 
\end{equation}
where $k$ is the ratio between the proton energy and the electron energy that we assumed to be 0 (i.e. purely leptonic), $\Phi$ is the filling factor that we assumed to be 1\footnote{The actual values of $k$ and $\Phi$ are poorly known but they do not affect the result in a significant way.}, $c_{12}$ is a constant tabulated in \citet{Pacholczyk_1970} that depends on the frequencies of observations and the spectral index $\alpha$, and $L$ and $V$ are respectively the luminosity and the volume of the emitting region, computed following Eq. \eqref{eq:LVeq}.
\begin{equation}
\label{eq:LVeq}
    L = \dfrac{4\pi D_L^2}{(1+z)^{1-\alpha}}\int^{\nu_2}_{\nu_1} F(\nu) \ ; \ V = \dfrac{4}{3} \pi \left(1.8\dfrac{\theta_2+\theta_1}{4} \dfrac{D_L}{(1+z)^2} \right)^3 
\end{equation}
where $D_L$ is the luminosity distance at redshift $z=0.048$, obtained through the Ned Wright's cosmological calculator \citep{Wright_2006}, and is equal to $\approx 214 \un{Mpc}$; the subscripts $1,2$ indicate, respectively, the lower and higher frequency taking in consideration; $F(\nu)$ is the flux density and $\theta$ is the size of the \texttt{modelfit} component. The various components are selected following the method outlined in \ref{sec:comprel}. 

We present an image of all 56 components found through this work in Fig. \ref{fig:allcomp}. The position of all the components is corrected for the core-shift found in \ref{sec:core_shift}.
From Fig. \ref{fig:allcomp}, it is possible to notice that some regions of the jet can be seen and modeled at various frequencies. This leads to a recognition criterion in order to identify said Gaussian components (see Sec. \ref{sec:comprel}).

\subsection{Polarization maps} \label{sec:polarization}
\begin{figure*}[h!]
    \centering
    \begin{subfigure}{0.45\textwidth}
        \includegraphics[width=\textwidth]{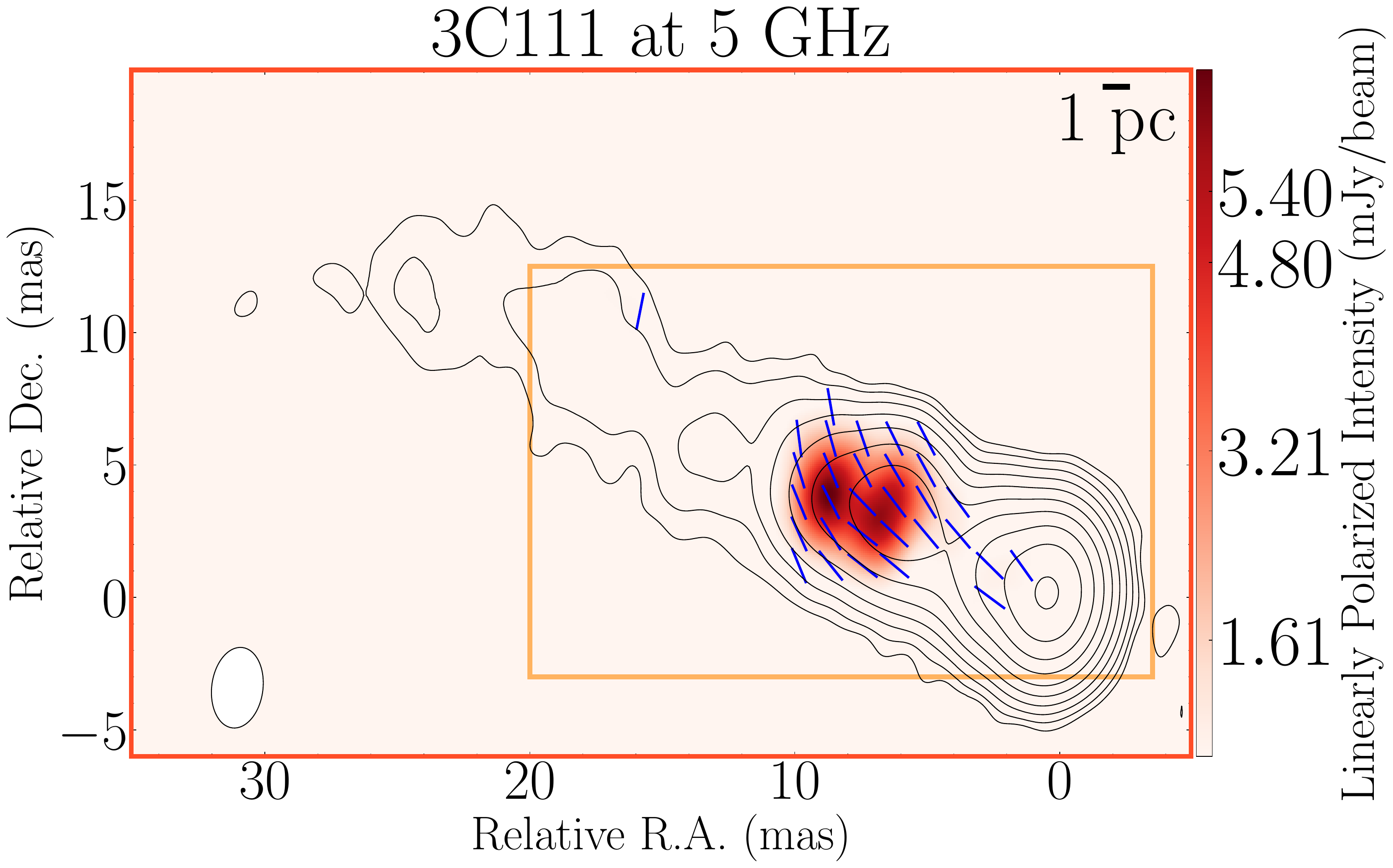}
        \caption{}
        \label{fig:polintc}
    \end{subfigure}
    \hfill
        \begin{subfigure}{0.45\textwidth}
        \includegraphics[width=\textwidth]{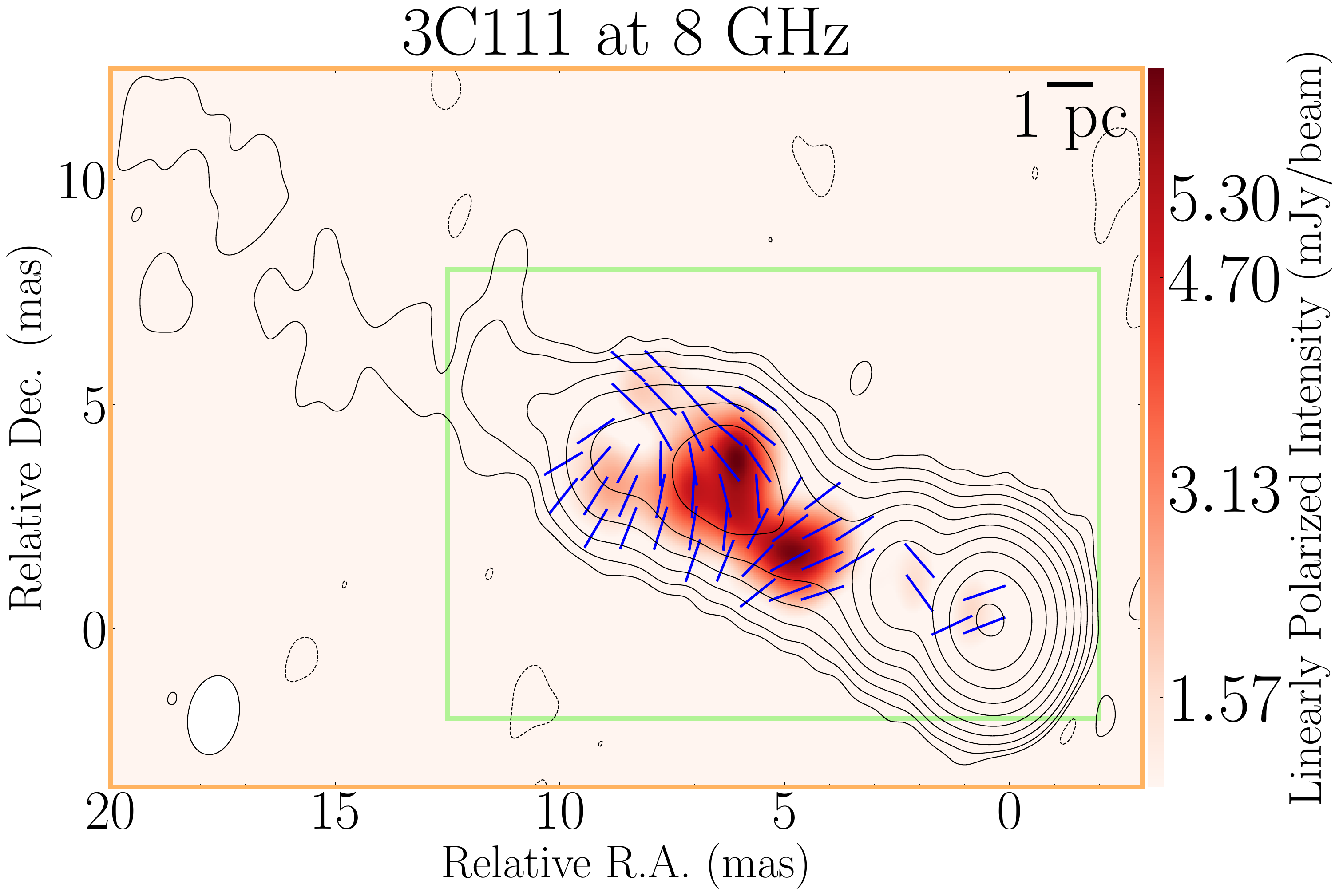}
        \caption{}
        \label{fig:polintx}
    \end{subfigure}
    \hfill
        \begin{subfigure}{0.45\textwidth}
        \includegraphics[width=\textwidth]{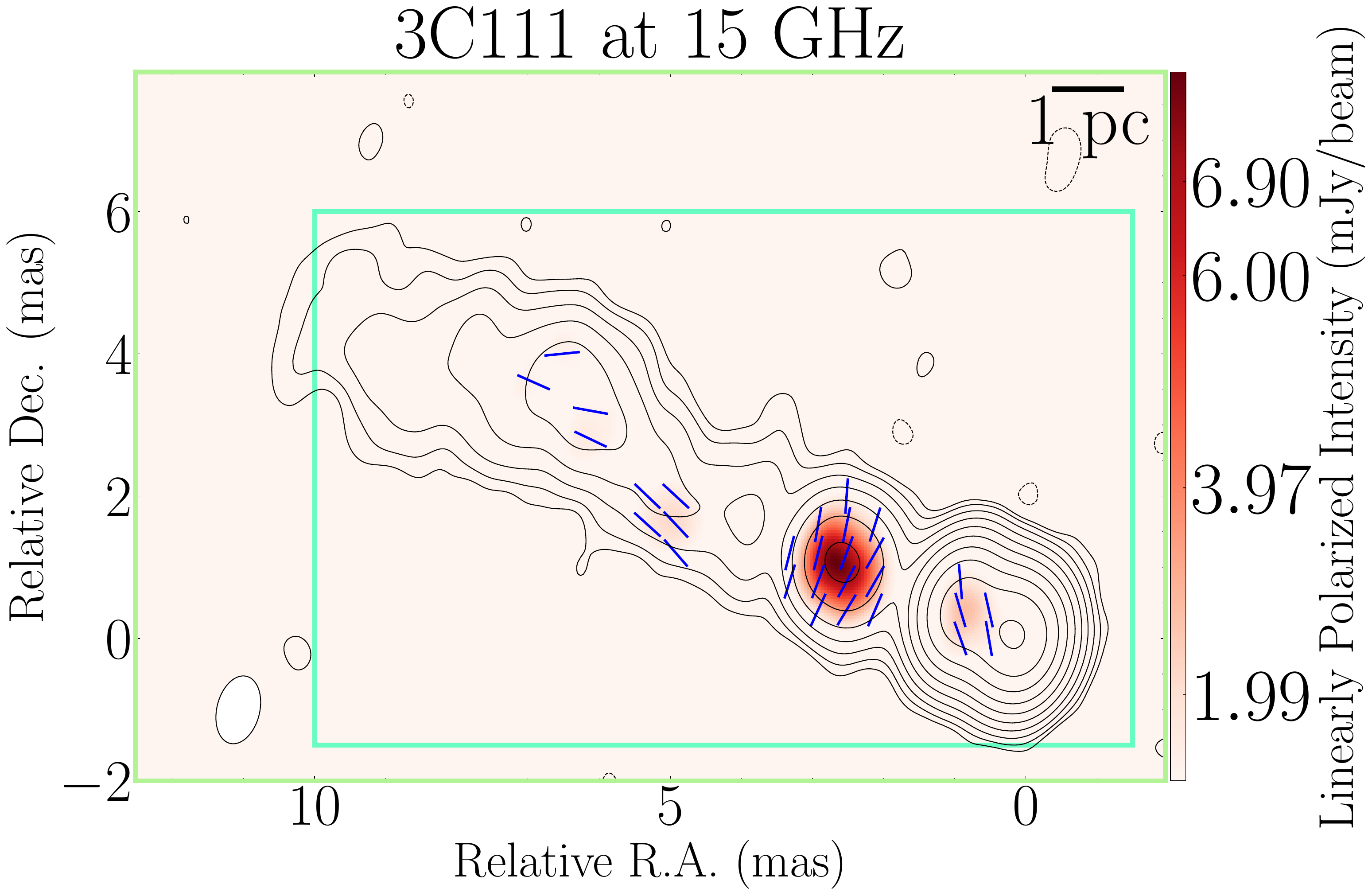}
        \caption{}
        \label{fig:polintu}
    \end{subfigure}
    \hfill
        \begin{subfigure}{0.45\textwidth}
        \includegraphics[width=\textwidth]{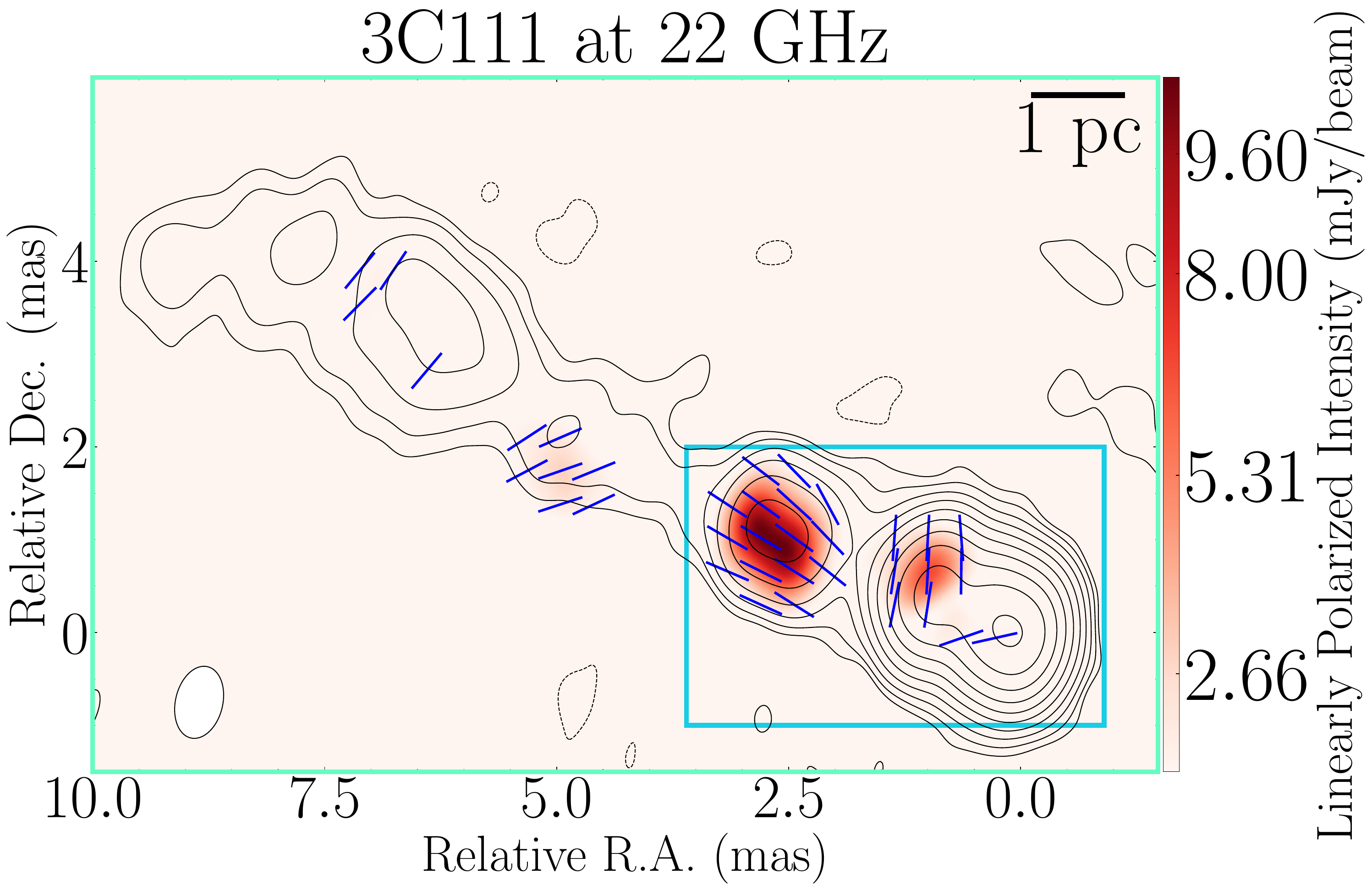}
        \caption{}
        \label{fig:polintk}
    \end{subfigure}
    \hfill
        \begin{subfigure}{0.45\textwidth}
        \includegraphics[width=\textwidth]{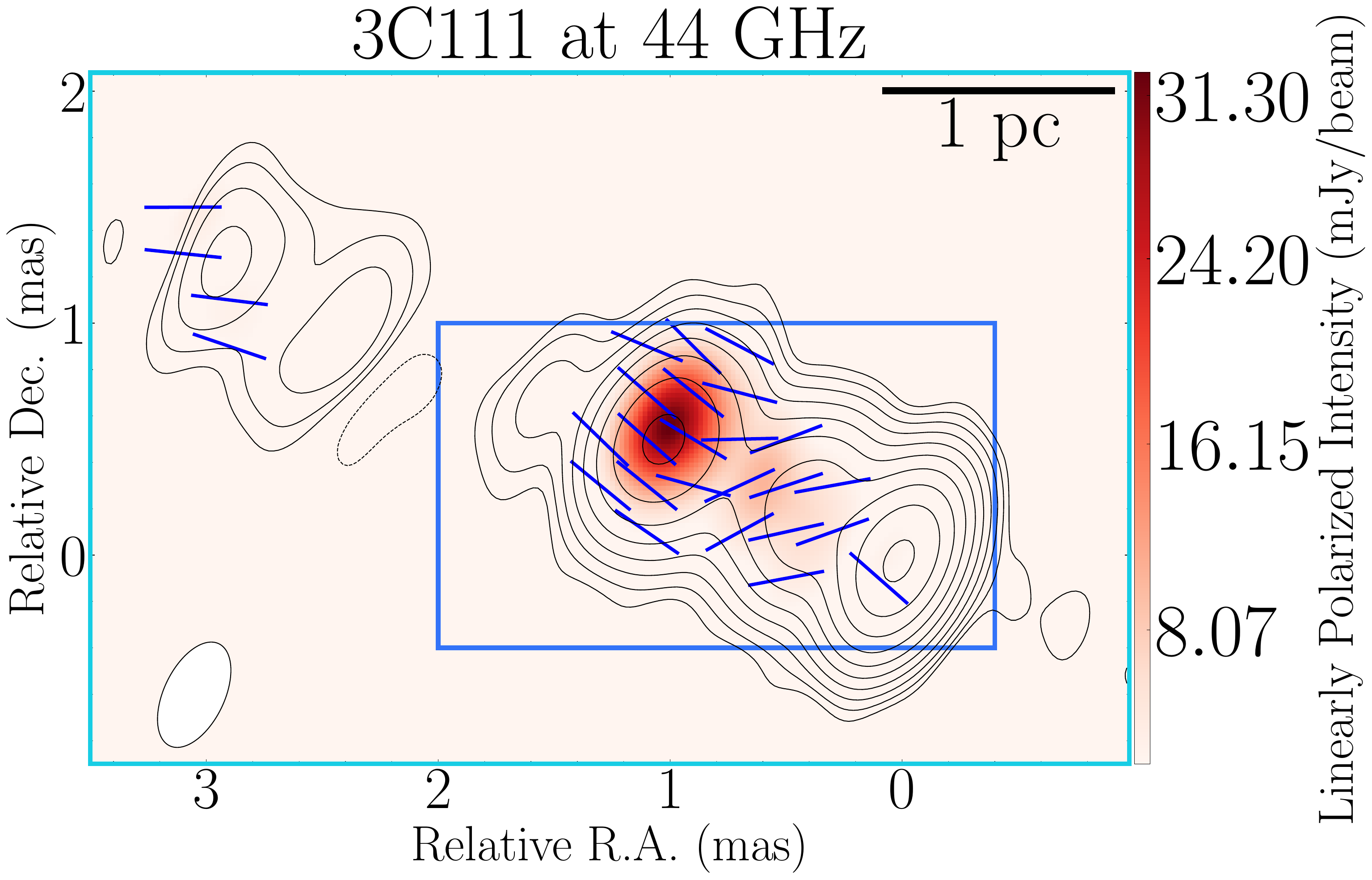}
        \caption{}
        \label{fig:polintq}
    \end{subfigure}
    \hfill
        \begin{subfigure}{0.45\textwidth}
        \includegraphics[width=\textwidth]{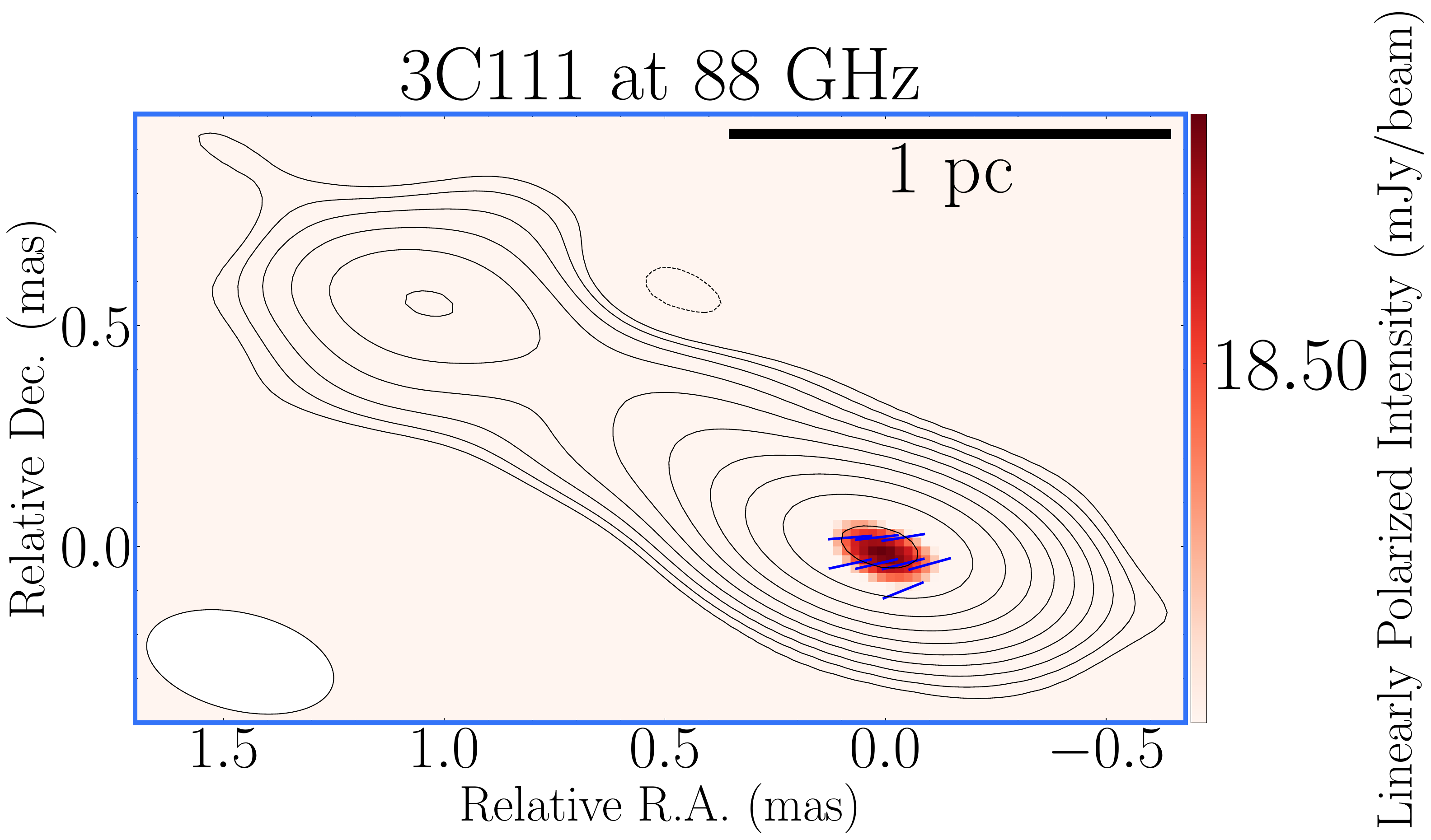}
        \caption{}
        \label{fig:polintw}
    \end{subfigure}
    \hfill
    \caption{Linear polarization intensity images of 3C\,111 for the 08/05/2014 observation, plotted over the respective total intensity contours (Fig. \ref{fig:totint}). In each panel is plotted the restoring beam as a white ellipse in the bottom left corner and a size scale reference of 1 pc in the top right corner. The blue bars represent the EVPAs. Starting from \ref{fig:polintc}, in each image, there is a colored box representing the size of the following frequency image. The frequencies are color-coded from the lowest $(5 \un{GHz})$ to the highest $(87.6  \un{GHz})$, going from red to blue. All the information for each map can be found in Tab. \ref{tab:polval}.}
    \label{fig:polint}
\end{figure*}
In order to retrieve information on the magnetic field in the jet of 3C\,111, we produced polarization maps for each frequency in the dataset.
We present the polarization maps at every available frequency (Fig. \ref{fig:polint}). In the final polarization maps, all pixels with linear polarization $m_l < 9 \sigma_{rms,pol}$ were blanked.
We computed $m_l$ both by integrating over the source and at the polarization peak $m_{l,peak}$. All the information is presented in Tab. \ref{tab:polval}. 
\begin{table*}
    \caption{\label{tab:polval} Total intensity and polarization data integrated over the whole structure detected in our observations.}   
    \centering
    \begin{tabular}{c c c c c c c c c c}
    \hline\hline
     $\nu$  & $F(\nu)$ & $F_P$ & $\sigma_\mathrm{rms,tot}$  & $P(\nu)$ & $P_P$ & $\sigma_\mathrm{rms,pol}$ & $m_\mathrm{l}$ & $m_\mathrm{l,peak}$ & EVPA \\
     \hline
        $5$ & $2930$ & $1800$ & $0.21$ & $16$ & $6$ & $0.07$ & $0.55 $ & $8.18 $ & $-33$ \\
        $8.4$ & $3840$ & $2280$ & $0.19$ & $30$ & $6$ & $0.10$ & $0.78 $ & $5.42 $ & $-14$\\
        $15.2$ & $3490 $ & $1460 $ & $0.23$ & $20 $ & $8 $ & $0.13$ & $0.57 $ & $6.35 $ & $53$ \\
        $21.9 \ $ & $3330 $ & $1310 $ & $0.46$ & $33 $ & $11 $ & $0.15$ & $0.99 $ & $14.87 $ & $49$ \\
        $43.8 \ $ & $2660 $ & $1310 $ & $1.14$ & $59 $ & $32 $ & $0.25$ & $2.21 $ & $9.09 $ & $37$ \\
        $87.6  \ $ & $1180 $ & $890 $ & $2.13$ & $5 $ & $19 $ & $1.90$ & $0.40 $ & $2.19 $ & $12$ \\
     \hline
    \end{tabular}
    \tablefoot{The columns report:
    (1) Frequency in $\un{GHz}$; (2) The total intensity flux density in $\un{mJy}$; (3) The peak in $\un{mJy/beam}$ and (4) the off-source rms in $\un{mJy/beam}$ in the total intensity full resolution image; (5) The total polarized flux in $\un{mJy}$; (6) The peak and (7) the off-source rms in $\un{mJy/beam}$ in the polarization image; (8) The linear fractional polarization in percentage; (9) The linear fractional polarization at the peak of the polarization image in percentage; (10) The calibrated EVPA computed in $\un{deg}$ following Sec.\ref{sec:Data_set}.}
    \label{tab:polval}    
\end{table*}
By looking at the total polarized flux of 3C\,111 at the highest frequency, it is possible to notice that this value is lower than the peak of the map. This is due to the different units and to the fact that at $87.6  \ \un{GHz}$, the region that has a polarization above the threshold is smaller than the beam size.

\subsection{Rotation Measure analysis} \label{sec:RotationMeasure}
When linearly polarized radiation crosses a magnetized medium with free electrons, the polarization angle $\chi$ (i.e., the EVPA) rotates from its intrinsic value $\chi_0$ following Eq. \eqref{eq:RMeq1} via a process called birefringence \citep[see][for a more complete review]{Beck_2015}:
\begin{equation}
\label{eq:RMeq1}
    \chi_{obs} = \chi_0 + RM \times \lambda^2 
\end{equation}
where $\lambda$ is the wavelength and $RM$ is the Rotation Measure, defined as Eq. \eqref{eq:RMeq2}:
\begin{equation}
\label{eq:RMeq2}  
\begin{split}
    RM & = \frac{e^3}{2 \pi m_e^2 c^4} \int^{d}_{0} n_e H_{\parallel} ds \\
    & = 8.1 \times 10^5 \int^{d}_{0} \left( \dfrac{n_e}{\un{cm^{-3}}} \right ) \left( \dfrac{H_{\parallel}}{\un{G}} \right ) \left( \dfrac{ds}{\un{pc}} \right ) \ \un{rad/m^2}
\end{split}
\end{equation}
where $n_e$ is the electron density, $H_{\parallel}$ is the magnetic field component of the intervening medium along the line of sight, and $d$ is the distance between the source and the observer.
The rotation may occur both internally to the radio-emitting region and externally, as long as a magnetized plasma is crossed by the polarized radiation anywhere along the line of sight.
By correcting the EVPAs for the effect of $RM$, it is also possible to study the intrinsic plane-of-sky or perpendicular electromagnetic field structure within the jet.
To produce the $RM$ maps, we combine two different sets of frequency triplets:  $5-8.4-15.2 \un{GHz}$ (Fig. \ref{fig:RM5815}) and $15.2-21.9-43.8 \un{GHz}$ (Fig. \ref{fig:RM152143}). 
Both the maps have been produced by first convolving the images at the three different frequencies with a common circular restoring beam of diameter 1.1 mas for the lower frequencies triplets and 0.4 mas for the higher ones. The images were then aligned with a cross-correlation script following an approach similar to Sec. \ref{sec:Spectral}.
To account for phase wrapping, we apply a correction to the EVPAs, making them flip by $180 \degree$ if their value combined with the respective error is $> 90 \degree$ or $< -90 \degree$.
We used the method of circular statistics outlined in \citet{Sarala_2001}, which fits $RM$ independently of the initial EVPA of the polarized emission. To account for the n$\pi$ ambiguity introduced by having a small number of frequencies for which to determine $RM$, we solve for n$\pi$ wraps of each EVPA, limiting our search for $RM$ up to $10^4 \frac{\text{rad}}{\text{m}^2}$. We applied the fitting function to each pixel with a linear polarization and Stokes I signal-to-noise ratio above $3\sigma$ and rejected any fits for which $\chi^2$ did not meet our significance threshold of 95\% (i.e., $\chi^2 > 3.84$) as done by \citet{Hovatta_2012}.
As \citet{Beuchert_2013} shows, for 3C111, the polarized emission at $5 \ \un{GHz}$ and $8.4 \ \un{GHz}$ can be ascribed to the core and the inner jet emission. Therefore, the integrated value of the EVPA can be used to calibrate the VLBA maps.
We can thus infer various insights from the $5-8.4-15.2 \un{GHz}$ map (Fig. \ref{fig:RM5815}). The region for which we manage to compute the $RM$ lies at $\approx 6 \ \un{pc}$ from the core.
The $15.2-21.9-43.8 \un{GHz}$ polarization images are sensitive to the inner portion of the jet, out to about 3 mas $(\approx 3 \un{pc})$ from the core. This allows us to study the $RM$ distribution close to the central engine (Fig. \ref{fig:RM152143}). 
The interpretation of these results is discussed in Sec. \ref{sec:polRM}.
\section{Discussion} \label{sec:Discussion}
In this section, we consider the various parameters measured in the previous one and search for correlations with the jet structure.
\subsection{Spectral index distribution} \label{sec:spixdis}
The spectral index analysis performed in Sec. \ref{sec:Spectral} can provide us with various insights on the electron population in the jet.
By looking at the spectral index distribution, we can infer that the jet base is optically thick ($\alpha > 0$) at all frequencies, becoming closer to a flat distribution ($\alpha = 0$) as the frequency rises. 
The jet emission stays optically thin throughout all frequencies, reaching values of $\approx -3$ in the $43.8-87.6  \un{GHz}$ map. 
The general $\alpha$ distribution is typical of AGN jets, and it is related to the optical depth of the emitting region \citep[see][for a complete review]{Urry_1995}.
Nonetheless, if we explore the spectral distribution in detail, we can find some discrepancies.
The jet shows some optically thick features at its edges in the three lowest frequency pairs. This can be explained by considering the low SNR and, consequently, the higher errors that are present at the edge of the jet brightness distribution. 
If we analyze the spectral index in the region at $\approx 6 \ \un{pc}$ from the center of the map, we can see a clear steepening in the value of $\alpha$ going from $-0.5$ in the $5-8.4 \ \un{GHz}$ map to $-2.5$ in the $15.2-21.9 \ \un{GHz}$ map. If intrinsic, this can be ascribed to multiple effects like radiative losses or multiple electron distributions in that region. The same effect has been found in other objects like M87 \citep{Nikonov_2023} or NGC\,315 \citep{Ricci_2022}.
A peculiar feature, that shows an inverted/flat distribution of $\alpha$ at all the frequency pairs but the highest one, is the one at $\approx 1.5 \ \un{mas}$ in R.A. and $\approx 0.5 \ \un{mas}$ in Dec. An inverted/flat spectrum in a region at $\approx 1.5 \un{pc}$ from the SMBH, does not have a unique interpretation but it could be associated with an external electron-dense region responsible for free-free absorption (see also Sec. \ref{sec:denseregion}).

\subsection{Modelfit components relations}\label{sec:comprel}
\begin{figure}
    \centering
    \includegraphics[width=\linewidth]{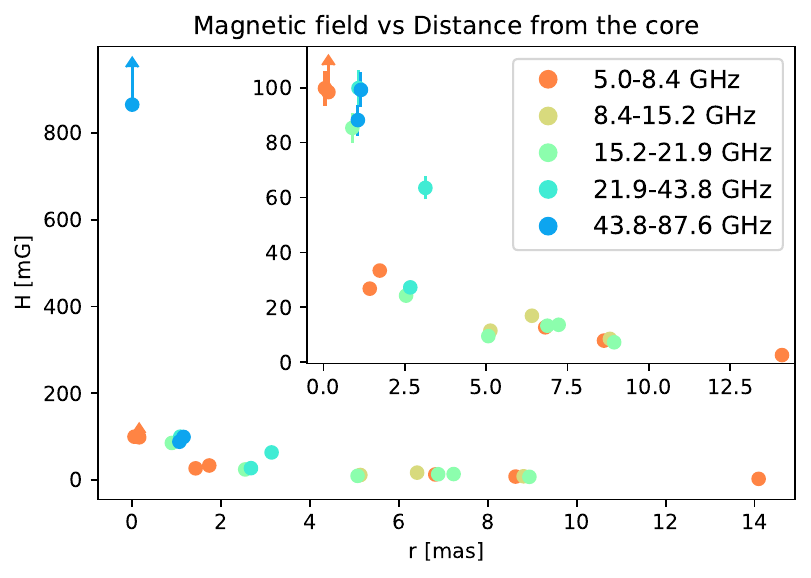}
    \caption{Equipartition magnetic field $H_{\mathrm{eq}}$ in $\un{mG}$ vs the distance from the core component $r$ in $\un{mas}$. The error bars represent the uncertainties, and the arrows represent the upper or lower limits. The various colors represent the different pairs of frequencies at which each component has been matched (\ref{sec:comprel}), as can be seen from the legend. The inset represents a zoomed-in version of the plot.}
    \label{fig:comprel}
\end{figure}
In this Section, we present an analysis of the physical quantities derived for the various components listed in Tab. \ref{tab:components}. As can be seen from Eq. \eqref{eq:LVeq}, we require components to be matched between multiple frequencies to determine $L$ and $V$.
We match components across frequencies based on their size $\theta$, position angle, and relative distance to the core $r$ (corrected for the core-shift effect), with the following conditions:
\begin{equation}\label{eq:comprec}
    ||\Vec{r_1} - \Vec{r_2}|| < (\theta_1/2 + \theta_2/2) \ \text{and} \ \dfrac{\theta_1}{\theta_2} < 1.3 \ \text{and} \ \dfrac{\theta_2}{\theta_1} < 0.7
\end{equation}
here, the subscript $1$ indicates the component at the lower frequency, and the subscript $2$ indicates the component at the higher one.
We retrieve a total of 22 components matched across at least two frequencies, as can be seen in Tab. \ref{tab:components_rec}. \\
For these components, we then investigate the interplay between the equipartition magnetic field  $H_{\mathrm{eq}}$  and the average value of different physical parameters. We notice that at the highest frequency pair (e.g. $87.6-43.8$ GHz), the core component passes our criteria because the difference in resolution is smaller than our threshold. We acknowledge this, and we interpret the result accordingly. 
In Fig. \ref{fig:comprel}, we show the equipartition magnetic field $H_{\mathrm{eq}}$ as a function of the distance from the core $r$. The magnetic field decreases as a function of distance with different trends based on the jet geometry \citep[e.g.][]{Boccardi_2021,Ricci_2022,Kovalev_2020}. The volume of the emitting region is increasing at larger $r$ while the flux density is decreasing, causing $H_{\mathrm{eq}}$ to decrease with the distance (Eq. \ref{eq:Heqeq}).
\begin{table*}
    \caption{\label{tab:components_rec} The 22 pairs of \texttt{modelfit} components selected following the criteria in Eq. \ref{eq:comprec}.}
    \centering
    \begin{tabular}{c c c c c}
    \hline
(1) $\nu_2-\nu_1 \un{(GHz)}$ & (2) $r_{mean} \un{(mas)}$ & (3) $\theta_{mean} \un{(mas)}$ & (4) $ \log(T_{mean}) \un{(K)}$ & (5) $H_{eq} \un{(mG)}$ \\
\hline
8.4 - 5.0 & $ 0.05 \pm 0.11 $ & $ 0.10 \pm 0.01 $ & $ 12.60 \pm 11.84 $ & $ 99.80 \pm 6.39 $\\ 

8.4 - 5.0 & $ 0.16 \pm 0.11 $ & $ < 0.10 $ & $ > 12.48 $ & $ > 98.39 $\\ 

8.4 - 5.0 & $ 1.43 \pm 0.11 $ & $ 0.26 \pm 0.02 $ & $ 11.23 \pm 10.55 $ & $ 26.81 \pm 1.73 $\\ 

8.4 - 5.0 & $ 1.73 \pm 0.11 $ & $ 0.23 \pm 0.02 $ & $ 11.31 \pm 10.57 $ & $ 33.40 \pm 2.14 $\\ 

8.4 - 5.0 & $ 6.82 \pm 0.11 $ & $ 0.56 \pm 0.04 $ & $ 10.36 \pm 9.67 $ & $ 12.71 \pm 0.81 $\\ 

8.4 - 5.0 & $ 8.62 \pm 0.11 $ & $ 0.81 \pm 0.06 $ & $ 9.69 \pm 8.95 $ & $ 7.88 \pm 0.51 $\\ 

8.4 - 5.0 & $ 14.09 \pm 0.11 $ & $ 1.49 \pm 0.11 $ & $ 8.31 \pm 7.59 $ & $ 2.61 \pm 0.17 $\\ 

15.2 - 8.4 & $ 5.13 \pm 0.06 $ & $ 0.50 \pm 0.04 $ & $ 9.48 \pm 8.78 $ & $ 11.50 \pm 0.74 $\\ 

15.2 - 8.4 & $ 6.41 \pm 0.06 $ & $ 0.38 \pm 0.03 $ & $ 9.95 \pm 9.25 $ & $ 16.94 \pm 1.08 $\\ 

15.2 - 8.4 & $ 8.80 \pm 0.06 $ & $ 0.77 \pm 0.05 $ & $ 9.11 \pm 8.37 $ & $ 8.52 \pm 0.55 $\\ 

21.9 - 15.2 & $ 0.89 \pm 0.04 $ & $ 0.12 \pm 0.01 $ & $ 11.48 \pm 10.73 $ & $ 85.36 \pm 5.48 $\\ 

21.9 - 15.2 & $ 2.54 \pm 0.04 $ & $ 0.26 \pm 0.02 $ & $ 9.90 \pm 9.13 $ & $ 24.26 \pm 1.55 $\\ 

21.9 - 15.2 & $ 5.07 \pm 0.04 $ & $ 0.49 \pm 0.03 $ & $ 8.74 \pm 7.97 $ & $ 9.52 \pm 0.62 $\\ 

21.9 - 15.2 & $ 6.88 \pm 0.04 $ & $ 0.44 \pm 0.03 $ & $ 9.16 \pm 8.40 $ & $ 13.30 \pm 0.87 $\\ 

21.9 - 15.2 & $ 7.23 \pm 0.04 $ & $ 0.42 \pm 0.03 $ & $ 9.22 \pm 8.46 $ & $ 13.65 \pm 0.89 $\\ 

21.9 - 15.2 & $ 8.93 \pm 0.04 $ & $ 0.75 \pm 0.05 $ & $ 8.56 \pm 7.81 $ & $ 7.27 \pm 0.47 $\\ 

43.8 - 21.9 & $ 1.08 \pm 0.03 $ & $ 0.12 \pm 0.01 $ & $ 10.94 \pm 10.23 $ & $ 99.86 \pm 6.41 $\\ 

43.8 - 21.9 & $ 2.67 \pm 0.03 $ & $ 0.27 \pm 0.02 $ & $ 9.40 \pm 8.71 $ & $ 27.27 \pm 1.75 $\\ 

43.8 - 21.9 & $ 3.14 \pm 0.03 $ & $ 0.08 \pm 0.01 $ & $ 10.16 \pm 9.48 $ & $ 63.51 \pm 4.17 $\\ 

\textbf{87.6 - 43.8} & $ 0.00 \pm 0.02 $ & $ < 0.01 $ & $ > 12.40 $ & $> 865.56 $\\ 

87.6 - 43.8 & $ 1.06 \pm 0.02 $ & $ 0.09 \pm 0.01 $ & $ 9.85 \pm 9.11 $ & $ 88.19 \pm 5.68 $\\ 

87.6 - 43.8 & $ 1.15 \pm 0.02 $ & $ 0.10 \pm 0.01 $ & $ 10.15 \pm 9.45 $ & $ 99.20 \pm 6.41 $\\ 
    \hline
    \end{tabular}
    \tablefoot{(1) Frequency pair in GHz (2) Mean distance and its error in mas (3) Mean angular size and its error in mas (4) Logarithm of the mean brightness temperature and its error in K (5) Equipartition magnetic field and its error in mG. The component in bold represents the core component at $87.6-43.8$ GHz.}
    \label{tab:components_rec}    
\end{table*}

\subsection{Polarized emission and Rotation Measure distribution} \label{sec:polRM}
\begin{figure*}
    \begin{subfigure}{0.5\textwidth}
        \includegraphics[width=\textwidth]{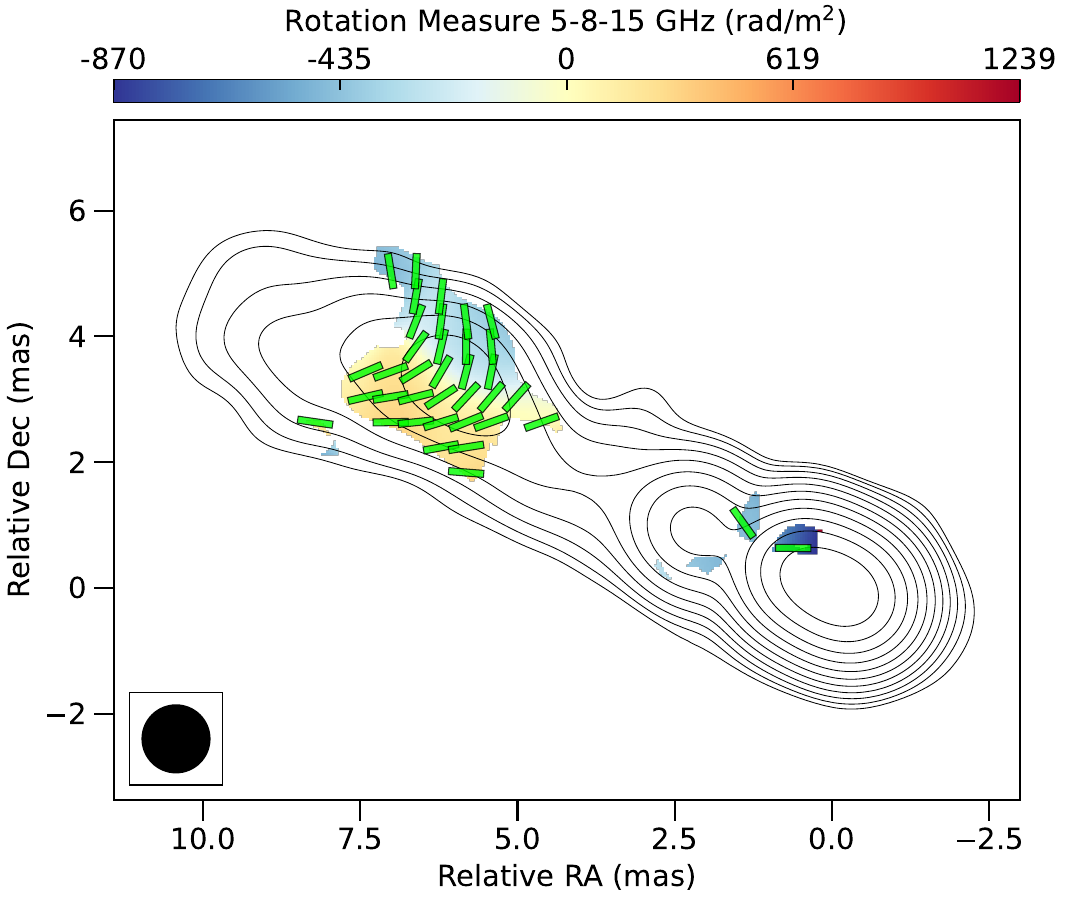}
        \caption{}
        \label{fig:RM5815}
    \end{subfigure}
    \hfill
    \begin{subfigure}{0.5\textwidth}
        \includegraphics[width=\textwidth]{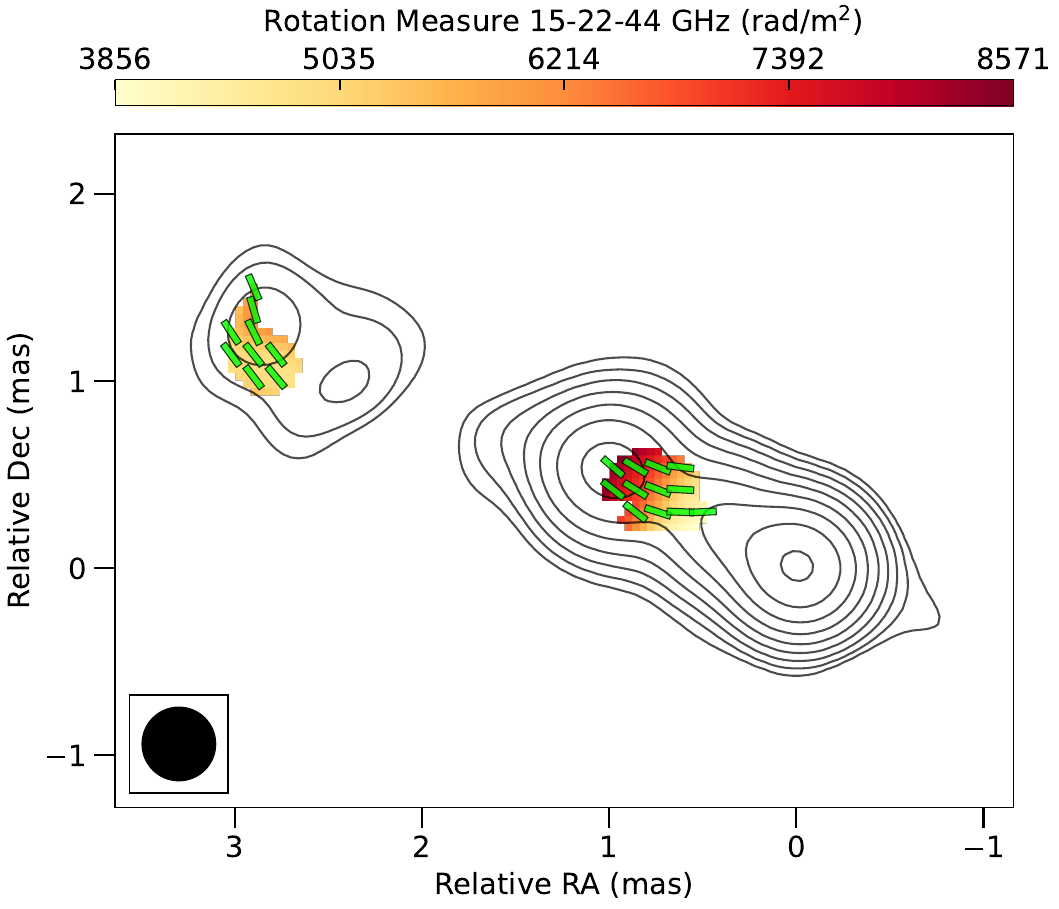}
        \caption{}
        \label{fig:RM152143}
    \end{subfigure}    
    \caption{\textbf{(a)}: $RM$ map between $5 \un{GHz}$, $8.4 \un{GHz}$, and $15.2 \un{GHz}$, plotted over the $15.2 \un{GHz}$ contours. The green bars correspond to the Faraday-corrected EVPAs. The $RM$ values are high because of the innermost region that likely yields more errors due to leakage or low polarization signal-to-noise. Moreover, this region has $\chi^2$ values close to our threshold, i.e., $3.84$.
    \textbf{(b)}: $RM$ map between $15.2 \un{GHz}$, $21.9 \un{GHz}$, and $43.8 \un{GHz}$, plotted over the $43.8 \un{GHz}$ contours. The green bars correspond to the Faraday-corrected EVPAs.}
    \label{fig:RM}
\end{figure*}
\begin{figure}
    \centering
    \includegraphics[width=\columnwidth]{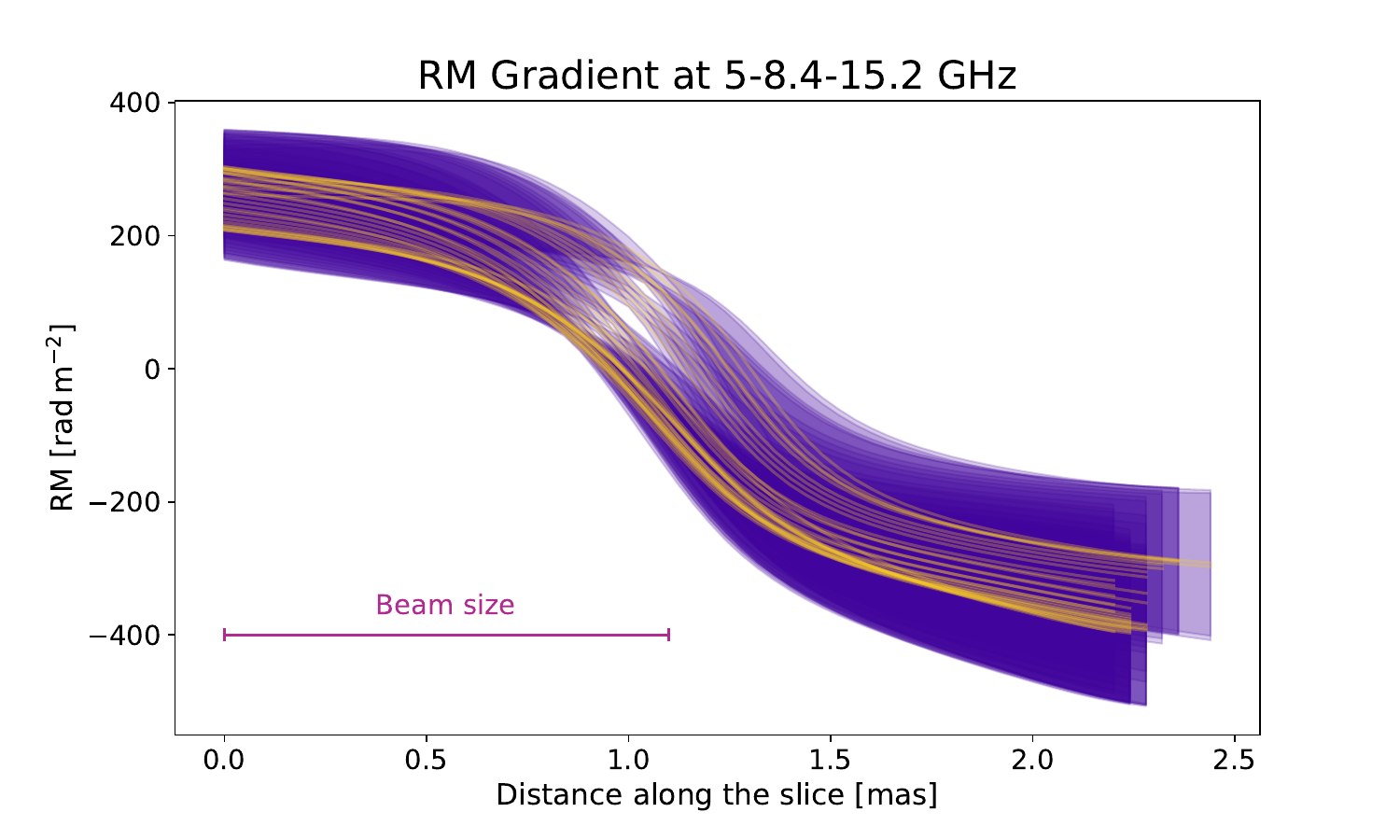}
    \caption{Transversal slices of $RM$ in the $5-8.4-15.2$ GHz map (Fig.\ref{fig:RM5815}) that are statistically significant (i.e. $> 2 \theta_{beam}$) \citep[see][for details]{Hovatta_2012}. The total number of slices is 33, each plotted with its relative error in purple. For reference, the beam size is plotted in the bottom left corner. The 33 slices occupy a region of $1.32\times 2.2 \ \un{mas}^2$.}
    \label{fig:rmgrad}
\end{figure}

As described in Sec. \ref{sec:polarization}, 3C\,111 shows polarized emission at all frequencies during the observation taken into account in this paper. 
The polarized emission lies mostly in the jet at all the frequencies, reaching its maximum value integrated over the whole emitting region at $43.8 \ \un{GHz}$ of $\approx 59 \ \un{mJy}$, corresponding to a polarization degree of $\approx 2\%$, integrated over the whole emitting region. As the wavelength decreases, this emission also arises from regions progressively closer to the jet base. At $87.6  \ \un{GHz}$ (Fig. \ref{fig:polintw}), we manage to retrieve polarized emission also from the core, with an integrated value over the whole emission region of $\approx 5 \ \un{mJy}$. By looking at Fig. \ref{fig:polint}, it is possible to recognize some features that are common at more than one frequency. For example, the polarized emission that arises at $43.8 \ \un{GHz}$ around $2 \ \un{pc}$ from the core (Fig. \ref{fig:polintq} can also be found in the lower frequency maps (e.g., Fig. \ref{fig:polintu}) with various intensities.
In Fig. \ref{fig:polintx}, we retrieve a polarization distribution similar to the one found in \citet{Beuchert_2018} in the stacked images at $15.2 \un{GHz}$. They found polarization in a smaller region slightly closer to the core.
Nonetheless, the EVPAs orientation is consistent with what they interpret as a continuously expanding and recollimating flow, thus supporting this scenario \citep[see][for more details]{Beuchert_2018}.\\
The combination of polarized emission at different frequencies allows us to study the $RM$ distribution in different jet regions, as seen in Fig. \ref{fig:RM}. We notice that the high values of the $RM$ are caused by the innermost jet emission that is more likely to have leakages between Stokes parameter, low polarization signal-to-noise, and it also has $\chi^2$ values close to our threshold, i.e. $3.84$. Therefore, we treat such regions as artifacts. 
As can be seen in Fig. \ref{fig:RM5815}, there seems to be a transversal gradient of $RM$ across the jet structure, going from about $-400 \un{rad/m^2}$ in the northern region to about $350 \un{rad/m^2}$ on the southern limb. The transition region between positive and negative values of $RM$ appears to lie in the spine of the jet. In Fig.\ref{fig:rmgrad} we show the $33$ slices that pass the criteria (with length ranging between $2-2.3 \times \theta_{beam}$ ) corresponding to a region of angular size $1.32 \times 2.2 \un{mas}^2$, thus confirming the significance of the spatial size of the gradient.
The median value of the maximum difference of $RM$ values across the transversal region is $611^{+171}_{-166} \un{rad}/\un{m}^2$. Such a significant $RM$ transversal gradient is strong evidence for the presence of a helical magnetic field in the jet region. \\
The magnetic field orientation is tilted by $\approx \pm 45 \degree$ with respect to the jet axis in the northern limb and southern limb, while it is parallel to the jet axis in the spine region. 
This distribution is in agreement with the simulated behavior of a helical magnetic field, as predicted by \citet{Fuentes_2021}. 
In Fig. \ref{fig:RM152143}, two regions can be seen: the brightest is at about 1 mas and shows $RM$ values from $\approx 3900$ to $\approx 8500 \ \un{rad/m^2}$. It is in between the two brightest components, visible in the total intensity emission. 
The magnetic field orientation is almost perpendicular to the jet axis in the outer part, and it is titled to $\approx 60 \degree$ with the jet axes in the inner part.
At a distance of about 3 mas, there is an additional region with $RM$ values $\approx 5000 \un{rad/m^2}$. Here, the magnetic field seems to be perpendicular to the jet axis.

\subsection{A possible denser region at $\approx$ 1.5 pc from the SMBH} \label{sec:denseregion}
From Fig. \ref{fig:spix}, it is possible to notice that at $\approx 1.5\un{pc}$ from the central engine, there is a region with a flat/inverted spectral index. In addition to that, the $RM$ map between $15.2-21.9-43.8 \ \un{GHz}$ (Fig. \ref{fig:RM152143}) shows a positive gradient going towards this region, possibly suggesting higher values of Faraday rotation. These two independent results are consistent with the presence of a region with high free electron density and/or strong magnetic fields. 
By using the virial relation Eq. \ref{eq:virial}, it is possible to have an estimate of the size of the Broad Line Region ($R_{BLR}$):
\begin{equation}
    R_{BLR} = \frac{M_{BH} G }{f (\Delta V)^2}
\label{eq:virial}
\end{equation}
where $M_{BH} = 10^{8.6} \un{M_{\odot}}$ is the Black Hole mass taken from \citet{Chatterjee_2011}, $G = 4.3 \times 10^{-3} \un{pc M_{\odot}^{-1} (km/s)^2}$ is the gravitational constant, $f$ is the virial factor, estimated with Eq. 6 in \citet{Yu_2019} and $\Delta V = 4800 \un{km/s}$ is the FWHM of the $H\alpha$ taken from \citet{Eracleous_2003}. From this estimate, the size of the BLR is $\approx 0.08 \un{pc}$, more than an order of magnitude smaller than the distance at which we observe a possible region with increased electron density. 
We, therefore, suggest the possibility that a cloud of the clumpy torus, which is thought to extend at $\approx 2 \un{pc}$ from the BH \citep{Cackett_2021}, is responsible for what we observe since it can explain both the flat/inverted spectrum at higher frequencies in a region different from the core and the high values of RM. 

\section{Conclusion} \label{sec:Conclusions}
In this paper, we present a set of multi-frequency, high angular resolution observations of pc-scale radio emission in the broad line radio galaxy 3C\,111. The images are obtained in full polarization with the VLBA, and as such, our work is the first study of this jet with simultaneous data from 5 GHz up to 87.6 GHz.
In the highest angular resolution image ($87.6  \un{GHz} $), our convolving beam is as small as $0.22 \ \un{mas}$, corresponding to $\approx 0.2$ pc.
The source is well detected at all frequencies, dominated by a compact core of brightness ranging between $\approx 3 \ \un{Jy}$ and $\approx 1 \ \un{Jy}$ (at $8.4 \ \un{GHz}$ and $87.6  \ \un{GHz}$, respectively). 
Our main results can be summarized as follows:
\begin{itemize}
    \item From the core, a one-sided jet emerges, extending up to $\approx 25 \ \un{mas}$ in length at $5 \ \un{GHz}$.  The jet is remarkably straight, being oriented at a position angle of $\approx 65 \degree$ and showing little to no evidence of bending over its full extension. Thanks to the multi-frequency analysis, it was possible to perform the core-shift analysis, where we found a trend of $r \propto \nu^{-1.27 \pm 0.19}$, implying $k_{\mathrm{r}} = 0.79 \pm 0.19$. This is suggestive of a slight dominance of the particle energy density with respect to the magnetic field.
    \item We computed the spectral index maps of the source. For the core, the spectral index varies from $\alpha \approx 1.5$ in the lower frequency pair $5-8.4 \ \un{GHz}$ to $\alpha \approx 0$ in the higher frequency pair $43.8-87.6  \ \un{GHz}$. The spectral index along the jet is generally steep, reaching the minimum value of $\alpha \approx -3$ in the highest frequency pair. When the same jet region is detected in more than one frequency pair, its spectral index steepens at a higher frequency. This can be due to different effects, like radiative losses or multi-layer electron distribution, as found in other AGN \citep[e.g.][]{Nikonov_2023,Ricci_2022}.
    \item From our model-fitting analysis, we found 56 components across all frequencies.
    We then estimated the brightness temperature distribution across the source for all the components. We found values in the range $10^7-10^{12}$ K. 
    Under the assumption that the emitting region is at the minimum total energy, the equipartition magnetic field was estimated for those components that we matched at different frequencies. We found values in the range of $1-100$ mG.
    \item The polarized emission lies mostly in the jet at all the frequencies and, with increasing frequency, this emission also arises from regions progressively closer to the jet base. The integrated fraction of polarization spans from a minimum of $0.40 \%$ at $87.6  \ \un{GHz}$, where only the core shows polarized emission, to $2.21 \%$ at $43.8 \ \un{GHz}$. Our results confirm the scenario proposed by \cite{Beuchert_2018} of an interaction of traveling features with a standing shock downstream in the jet.
    \item By considering the polarized regions common to various frequencies, we computed the $RM$ for two triplets of frequencies.
    For the $15.2-21.9-43.8 \ \un{GHz}$ triplet, we identified two regions: the brightest one is at $\approx 2 \ \un{pc}$ from the core and shows a gradient of $RM$ from $\approx 3900$ to $\approx 8400 \ \un{rad/m^2}$; at a distance of about $3 \ \un{mas}$, lies another region with $RM$ values of $\approx 6000 \ \un{rad/m^2}$. In this outer region, the magnetic field seems to start aligning with the jet axis. At the lowest frequencies, the region with common polarised emission is at a distance of $\approx 6 \un{mas}$ from the core. In this region, there is a significant transverse gradient of $RM$, going from about $250 \un{rad/m^2}$ down to $\approx -400 \un{rad/m^2}$. The magnetic field orientation is perpendicular to the jet axis in the central region, where the transition from positive to negative values of $RM$ happens. In the outer region, the magnetic field seems to align with the jet axis. This magnetic field distribution is in agreement with a helical structure \citep{Fuentes_2021}. 
    \item By looking at the spectral index map between $21.9 \ \un{GHz}$ and $43.8 \ \un{GHz}$ and the $RM$ between $15.2-21.9-43.8 \ \un{GHz}$, there seems to be a co-spatiality between the region of the flat/inverted spectrum and the region with high value of $RM$. We propose, as a possible explanation, a cloud of the clumpy torus that lies at $\approx 1.5\ \un{pc}$ from the SMBH.
\end{itemize}

\begin{acknowledgements} 
We would like to thank Dr. M. Kadler and Dr. B. Boccardi for their helpful discussion, which improved the manuscript.
This paper made use of "Uncertainties: a Python package for calculations with uncertainties, Eric O. LEBIGOT", http://pythonhosted.org/uncertainties/.
This research has made use of data from the MOJAVE database that is maintained by the MOJAVE team \citep{Lister_2018}. 
This research has been partially funded by the Deutsche Forschungsgemeinschaft (DFG, German Research Foundation) – project number 443220636 (DFG research unit FOR 5195: "Relativistic Jets in Active Galaxies"). 
J.D.L. would like to acknowledge that this publication is part of the M2FINDERS project, which has received funding from the European Research Council (ERC) under the European Union’s Horizon 2020 Research and Innovation Programme (grant agreement No 101018682).
\end{acknowledgements}

\bibliographystyle{aa}
\bibliography{bibliography}

\let\cleardoublepage\clearpage

\begin{appendix}
\onecolumn 
\section{Modelfit components}
\begin{table}[!ht]
    \tiny
    \begin{minipage}{\textwidth}
    \caption{\label{tab:components} All 56 \texttt{modelfit} components parameters with respective uncertainties (see Sec. \ref{sec:Modelfit}).}
    \centering
    \begin{tabular}{c c c c c c c c c c}
    \hline\hline
ID & $\nu$ (GHz) & $F$ (Jy) & $r$ (mas) & $\phi$ (deg) & $r_{cs}$ (mas) & $\phi_{cs}$ (deg) & $\theta$ (mas) & $T$ (K) \\ 
\hline
\textbf{C0} & 5.0 & $ 1.23  \pm  0.12 $ & $ 0.29  \pm  0.19 $ & $ 80.6  \pm  8.1 $ & $ 0.00  \pm  0.19 $ & $ 0.0  \pm  8.1 $ & $ 0.10  \pm  0.01 $ & $ (1.5 \pm 0.3) \times 10^{12} $ \\ 

C1 & 5.0 & $ 0.68  \pm  0.07 $ & $ 0.43  \pm  0.19 $ & $ 55.5  \pm  5.5 $ & $ 0.21  \pm  0.19 $ & $ 19.8  \pm  5.5 $ & $ < 0.10 $ & $ > 9.8 \times 10^{11} $ \\ 

C2 & 5.0 & $ 0.33  \pm  0.03 $ & $ 1.81  \pm  0.19 $ & $ 70.1  \pm  7.0 $ & $ 1.52  \pm  0.19 $ & $ 68.1  \pm  7.0 $ & $ 0.23  \pm  0.02 $ & $ (8.0 \pm 1.8) \times 10^{10} $ \\ 

C3 & 5.0 & $ 0.14  \pm  0.01 $ & $ 2.57  \pm  0.19 $ & $ 68.8  \pm  6.9 $ & $ 2.28  \pm  0.19 $ & $ 67.3  \pm  6.9 $ & $ 0.20  \pm  0.02 $ & $ (4.8 \pm 1.1) \times 10^{10} $ \\ 

C4 & 5.0 & $ 0.08  \pm  0.01 $ & $ 4.39  \pm  0.19 $ & $ 68.3  \pm  6.8 $ & $ 4.11  \pm  0.19 $ & $ 67.5  \pm  6.8 $ & $ 0.46  \pm  0.05 $ & $ (4.6 \pm 1.0) \times 10^{9} $ \\ 

C5 & 5.0 & $ 0.09  \pm  0.01 $ & $ 5.81  \pm  0.19 $ & $ 63.0  \pm  6.3 $ & $ 5.53  \pm  0.19 $ & $ 62.1  \pm  6.3 $ & $ 0.25  \pm  0.03 $ & $ (1.8 \pm 0.4) \times 10^{10} $ \\ 

C6 & 5.0 & $ 0.23  \pm  0.02 $ & $ 6.82  \pm  0.19 $ & $ 62.1  \pm  6.2 $ & $ 6.54  \pm  0.19 $ & $ 61.2  \pm  6.2 $ & $ 0.53  \pm  0.05 $ & $ (1.0 \pm 0.2) \times 10^{10} $ \\ 

C7 & 5.0 & $ 0.12  \pm  0.01 $ & $ 8.70  \pm  0.19 $ & $ 65.6  \pm  6.6 $ & $ 8.42  \pm  0.19 $ & $ 65.1  \pm  6.6 $ & $ 0.88  \pm  0.09 $ & $ (1.9 \pm 0.4) \times 10^{9} $ \\ 

C8 & 5.0 & $ 0.02  \pm  0.00 $ & $ 14.38  \pm  0.19 $ & $ 65.8  \pm  6.6 $ & $ 14.10  \pm  0.19 $ & $ 65.5  \pm  6.6 $ & $ 1.57  \pm  0.16 $ & $ (8.4 \pm 1.9) \times 10^{7} $ \\ 

C9 & 5.0 & $ 0.01  \pm  0.00 $ & $ 23.07  \pm  0.19 $ & $ 62.2  \pm  6.2 $ & $ 22.79  \pm  0.19 $ & $ 61.9  \pm  6.2 $ & $ 2.73  \pm  0.27 $ & $ (2.3 \pm 0.5) \times 10^{7} $ \\ 

\textbf{X0} & 8.4 & $ 0.97  \pm  0.10 $ & $ 0.20  \pm  0.11 $ & $ 65.5  \pm  6.5 $ & $ 0.00  \pm  0.11 $ & $ 0.0  \pm  6.5 $ & $ 0.08  \pm  0.01 $ & $ (7.3 \pm 1.6) \times 10^{11} $ \\ 

X1 & 8.4 & $ 1.27  \pm  0.13 $ & $ 0.30  \pm  0.11 $ & $ 68.9  \pm  6.9 $ & $ 0.10  \pm  0.11 $ & $ 75.5  \pm  6.9 $ & $ 0.10  \pm  0.01 $ & $ (5.4 \pm 1.2) \times 10^{11} $ \\ 

X2 & 8.4 & $ 0.81  \pm  0.08 $ & $ 0.65  \pm  0.11 $ & $ 62.9  \pm  6.3 $ & $ 0.45  \pm  0.11 $ & $ 61.8  \pm  6.3 $ & $ < 0.06 $ & $ > 1.2 \times 10^{12} $ \\ 

X3 & 8.4 & $ 0.10  \pm  0.01 $ & $ 1.53  \pm  0.11 $ & $ 69.0  \pm  6.9 $ & $ 1.33  \pm  0.11 $ & $ 69.5  \pm  6.9 $ & $ 0.28  \pm  0.03 $ & $ (5.9 \pm 1.3) \times 10^{9} $ \\ 

X4 & 8.4 & $ 0.26  \pm  0.03 $ & $ 2.14  \pm  0.11 $ & $ 67.8  \pm  6.8 $ & $ 1.94  \pm  0.11 $ & $ 68.1  \pm  6.8 $ & $ 0.22  \pm  0.02 $ & $ (2.4 \pm 0.5) \times 10^{10} $ \\ 

X5 & 8.4 & $ 0.05  \pm  0.01 $ & $ 3.90  \pm  0.11 $ & $ 67.9  \pm  6.8 $ & $ 3.70  \pm  0.11 $ & $ 68.0  \pm  6.8 $ & $ 0.60  \pm  0.06 $ & $ (6.9 \pm 1.5) \times 10^{8} $ \\ 

X6 & 8.4 & $ 0.07  \pm  0.01 $ & $ 5.38  \pm  0.11 $ & $ 64.7  \pm  6.5 $ & $ 5.18  \pm  0.11 $ & $ 64.7  \pm  6.5 $ & $ 0.47  \pm  0.05 $ & $ (1.3 \pm 0.3) \times 10^{9} $ \\ 

X7 & 8.4 & $ 0.12  \pm  0.01 $ & $ 6.44  \pm  0.11 $ & $ 60.7  \pm  6.1 $ & $ 6.24  \pm  0.11 $ & $ 60.6  \pm  6.1 $ & $ 0.38  \pm  0.04 $ & $ (3.9 \pm 0.9) \times 10^{9} $ \\ 

X8 & 8.4 & $ 0.09  \pm  0.01 $ & $ 7.29  \pm  0.11 $ & $ 62.7  \pm  6.3 $ & $ 7.09  \pm  0.11 $ & $ 62.6  \pm  6.3 $ & $ 0.58  \pm  0.06 $ & $ (1.2 \pm 0.3) \times 10^{9} $ \\ 

X9 & 8.4 & $ 0.06  \pm  0.01 $ & $ 9.02  \pm  0.11 $ & $ 65.7  \pm  6.6 $ & $ 8.82  \pm  0.11 $ & $ 65.7  \pm  6.6 $ & $ 0.75  \pm  0.07 $ & $ (5.1 \pm 1.1) \times 10^{8} $ \\ 

X10 & 8.4 & $ 0.01  \pm  0.00 $ & $ 14.28  \pm  0.11 $ & $ 64.9  \pm  6.5 $ & $ 14.08  \pm  0.11 $ & $ 64.9  \pm  6.5 $ & $ 1.41  \pm  0.14 $ & $ (1.7 \pm 0.4) \times 10^{7} $ \\ 

\textbf{U0} & 15.2 & $ 0.88  \pm  0.09 $ & $ 0.02  \pm  0.06 $ & $ 172.6  \pm  17.3 $ & $ 0.00  \pm  0.06 $ & $ 0.0  \pm  17.3 $ & $ < 0.03 $ & $ > 1.3 \times 10^{12} $ \\ 

U1 & 15.2 & $ 0.49  \pm  0.05 $ & $ 0.29  \pm  0.06 $ & $ 58.0  \pm  5.8 $ & $ 0.30  \pm  0.06 $ & $ 54.5  \pm  5.8 $ & $ < 0.03 $ & $ > 7.5 \times 10^{11} $ \\ 

U2 & 15.2 & $ 0.56  \pm  0.06 $ & $ 0.30  \pm  0.06 $ & $ 63.1  \pm  6.3 $ & $ 0.30  \pm  0.06 $ & $ 59.5  \pm  6.3 $ & $ 0.06  \pm  0.01 $ & $ (1.9 \pm 0.4) \times 10^{11} $ \\ 

U3 & 15.2 & $ 0.99  \pm  0.10 $ & $ 0.78  \pm  0.06 $ & $ 64.9  \pm  6.5 $ & $ 0.79  \pm  0.06 $ & $ 63.5  \pm  6.5 $ & $ 0.11  \pm  0.01 $ & $ (1.1 \pm 0.3) \times 10^{11} $ \\ 

U4 & 15.2 & $ 0.11  \pm  0.01 $ & $ 1.08  \pm  0.06 $ & $ 66.1  \pm  6.6 $ & $ 1.09  \pm  0.06 $ & $ 65.1  \pm  6.6 $ & $ 0.05  \pm  0.01 $ & $ (5.8 \pm 1.3) \times 10^{10} $ \\ 

U5 & 15.2 & $ 0.14  \pm  0.01 $ & $ 2.45  \pm  0.06 $ & $ 68.4  \pm  6.8 $ & $ 2.45  \pm  0.06 $ & $ 67.9  \pm  6.8 $ & $ 0.26  \pm  0.03 $ & $ (2.8 \pm 0.6) \times 10^{9} $ \\ 

U6 & 15.2 & $ 0.08  \pm  0.01 $ & $ 2.86  \pm  0.06 $ & $ 66.1  \pm  6.6 $ & $ 2.87  \pm  0.06 $ & $ 65.7  \pm  6.6 $ & $ 0.17  \pm  0.02 $ & $ (3.7 \pm 0.8) \times 10^{9} $ \\ 

U7 & 15.2 & $ 0.04  \pm  0.00 $ & $ 5.08  \pm  0.06 $ & $ 67.6  \pm  6.8 $ & $ 5.08  \pm  0.06 $ & $ 67.3  \pm  6.8 $ & $ 0.53  \pm  0.05 $ & $ (1.9 \pm 0.4) \times 10^{8} $ \\ 

U8 & 15.2 & $ 0.05  \pm  0.01 $ & $ 6.57  \pm  0.06 $ & $ 63.0  \pm  6.3 $ & $ 6.57  \pm  0.06 $ & $ 62.8  \pm  6.3 $ & $ 0.37  \pm  0.04 $ & $ (5.1 \pm 1.2) \times 10^{8} $ \\ 

U9 & 15.2 & $ 0.06  \pm  0.01 $ & $ 7.26  \pm  0.06 $ & $ 60.4  \pm  6.0 $ & $ 7.27  \pm  0.06 $ & $ 60.3  \pm  6.0 $ & $ 0.35  \pm  0.04 $ & $ (6.1 \pm 1.4) \times 10^{8} $ \\ 

U10 & 15.2 & $ 0.06  \pm  0.01 $ & $ 8.78  \pm  0.06 $ & $ 64.5  \pm  6.4 $ & $ 8.79  \pm  0.06 $ & $ 64.4  \pm  6.4 $ & $ 0.79  \pm  0.08 $ & $ (1.4 \pm 0.3) \times 10^{8} $ \\ 

\textbf{K0} & 21.9 & $ 0.69  \pm  0.07 $ & $ 0.04  \pm  0.05 $ & $ 116.4  \pm  11.6 $ & $ 0.00  \pm  0.05 $ & $ 0.0  \pm  11.6 $ & $ 0.08  \pm  0.01 $ & $ (7.6 \pm 1.7) \times 10^{10} $ \\ 

K1 & 21.9 & $ 0.38  \pm  0.04 $ & $ 0.14  \pm  0.05 $ & $ 45.2  \pm  4.5 $ & $ 0.13  \pm  0.05 $ & $ 26.8  \pm  4.5 $ & $ 0.06  \pm  0.01 $ & $ (7.4 \pm 1.7) \times 10^{10} $ \\ 

K2 & 21.9 & $ 0.51  \pm  0.05 $ & $ 0.28  \pm  0.05 $ & $ 71.7  \pm  7.2 $ & $ 0.26  \pm  0.05 $ & $ 64.9  \pm  7.2 $ & $ 0.05  \pm  0.00 $ & $ (1.6 \pm 0.4) \times 10^{11} $ \\ 

K3 & 21.9 & $ 0.47  \pm  0.05 $ & $ 0.49  \pm  0.05 $ & $ 62.4  \pm  6.2 $ & $ 0.47  \pm  0.05 $ & $ 58.2  \pm  6.2 $ & $ 0.14  \pm  0.01 $ & $ (1.6 \pm 0.4) \times 10^{10} $ \\ 

K4 & 21.9 & $ 0.91  \pm  0.09 $ & $ 1.02  \pm  0.05 $ & $ 64.9  \pm  6.5 $ & $ 0.99  \pm  0.05 $ & $ 63.0  \pm  6.5 $ & $ 0.13  \pm  0.01 $ & $ (3.7 \pm 0.8) \times 10^{10} $ \\ 

K5 & 21.9 & $ 0.04  \pm  0.00 $ & $ 1.43  \pm  0.05 $ & $ 67.0  \pm  6.7 $ & $ 1.41  \pm  0.05 $ & $ 65.7  \pm  6.7 $ & $ 0.10  \pm  0.01 $ & $ (2.8 \pm 0.6) \times 10^{9} $ \\ 

K6 & 21.9 & $ 0.12  \pm  0.01 $ & $ 2.66  \pm  0.05 $ & $ 67.3  \pm  6.7 $ & $ 2.63  \pm  0.05 $ & $ 66.6  \pm  6.7 $ & $ 0.26  \pm  0.03 $ & $ (1.1 \pm 0.3) \times 10^{9} $ \\ 

K7 & 21.9 & $ 0.05  \pm  0.01 $ & $ 3.10  \pm  0.05 $ & $ 67.0  \pm  6.7 $ & $ 3.07  \pm  0.05 $ & $ 66.4  \pm  6.7 $ & $ 0.07  \pm  0.01 $ & $ (6.6 \pm 1.5) \times 10^{9} $ \\ 

K8 & 21.9 & $ 0.03  \pm  0.00 $ & $ 5.08  \pm  0.05 $ & $ 66.6  \pm  6.7 $ & $ 5.06  \pm  0.05 $ & $ 66.3  \pm  6.7 $ & $ 0.45  \pm  0.04 $ & $ (8.6 \pm 2.0) \times 10^{7} $ \\ 

K9 & 21.9 & $ 0.08  \pm  0.01 $ & $ 7.22  \pm  0.05 $ & $ 62.2  \pm  6.2 $ & $ 7.19  \pm  0.05 $ & $ 61.9  \pm  6.2 $ & $ 0.50  \pm  0.05 $ & $ (2.1 \pm 0.5) \times 10^{8} $ \\ 

K10 & 21.9 & $ 0.03  \pm  0.00 $ & $ 9.11  \pm  0.05 $ & $ 63.3  \pm  6.3 $ & $ 9.08  \pm  0.05 $ & $ 63.1  \pm  6.3 $ & $ 0.72  \pm  0.07 $ & $ (4.4 \pm 1.0) \times 10^{7} $ \\ 

\textbf{Q0} & 43.8 & $ 0.98  \pm  0.10 $ & $ 0.02  \pm  0.03 $ & $ 102.5  \pm  10.3 $ & $ 0.00  \pm  0.03 $ & $ 0.0  \pm  10.3 $ & $ < 0.01 $ & $ > 9.7 \times 10^{11} $ \\ 

Q1 & 43.8 & $ 0.29  \pm  0.03 $ & $ 0.08  \pm  0.03 $ & $ 76.6  \pm  7.7 $ & $ 0.07  \pm  0.03 $ & $ 70.8  \pm  7.7 $ & $ 0.03  \pm  0.00 $ & $ (4.2 \pm 0.9) \times 10^{10} $ \\ 

Q2 & 43.8 & $ 0.25  \pm  0.02 $ & $ 0.14  \pm  0.03 $ & $ 58.0  \pm  5.8 $ & $ 0.13  \pm  0.03 $ & $ 52.9  \pm  5.8 $ & $ < 0.01 $ & $ > 2.4 \times 10^{11} $ \\ 

Q3 & 43.8 & $ 0.28  \pm  0.03 $ & $ 0.39  \pm  0.03 $ & $ 61.9  \pm  6.2 $ & $ 0.38  \pm  0.03 $ & $ 60.3  \pm  6.2 $ & $ 0.05  \pm  0.01 $ & $ (1.8 \pm 0.4) \times 10^{10} $ \\ 

Q4 & 43.8 & $ 0.19  \pm  0.02 $ & $ 0.61  \pm  0.03 $ & $ 57.8  \pm  5.8 $ & $ 0.59  \pm  0.03 $ & $ 56.7  \pm  5.8 $ & $ 0.10  \pm  0.01 $ & $ (3.4 \pm 0.8) \times 10^{9} $ \\ 

Q5 & 43.8 & $ 0.16  \pm  0.02 $ & $ 0.99  \pm  0.03 $ & $ 62.7  \pm  6.3 $ & $ 0.98  \pm  0.03 $ & $ 62.1  \pm  6.3 $ & $ 0.10  \pm  0.01 $ & $ (2.8 \pm 0.6) \times 10^{9} $ \\ 

Q6 & 43.8 & $ 0.42  \pm  0.04 $ & $ 1.18  \pm  0.03 $ & $ 62.1  \pm  6.2 $ & $ 1.17  \pm  0.03 $ & $ 61.6  \pm  6.2 $ & $ 0.11  \pm  0.01 $ & $ (6.3 \pm 1.4) \times 10^{9} $ \\ 

Q7 & 43.8 & $ 0.05  \pm  0.01 $ & $ 2.73  \pm  0.03 $ & $ 66.5  \pm  6.6 $ & $ 2.71  \pm  0.03 $ & $ 66.3  \pm  6.6 $ & $ 0.27  \pm  0.03 $ & $ (1.2 \pm 0.3) \times 10^{8} $ \\ 

Q8 & 43.8 & $ 0.03  \pm  0.00 $ & $ 3.21  \pm  0.03 $ & $ 65.6  \pm  6.6 $ & $ 3.20  \pm  0.03 $ & $ 65.4  \pm  6.6 $ & $ 0.08  \pm  0.01 $ & $ (6.7 \pm 1.7) \times 10^{8} $ \\ 

\textbf{W0} & 87.6 & $ 0.87  \pm  0.09 $ & $ 0.02  \pm  0.02 $ & $ 17.5  \pm  1.7 $ & $ 0.00  \pm  0.02 $ & $ 0.0  \pm  1.7 $ & $ < 0.01 $ & $ > 3.0 \times 10^{11} $ \\ 

W1 & 87.6 & $ 0.08  \pm  0.01 $ & $ 0.25  \pm  0.02 $ & $ 68.5  \pm  6.8 $ & $ 0.24  \pm  0.02 $ & $ 71.5  \pm  6.8 $ & $ < 0.01 $ & $ > 2.7 \times 10^{10} $ \\ 

W2 & 87.6 & $ 0.10  \pm  0.01 $ & $ 0.50  \pm  0.02 $ & $ 61.5  \pm  6.1 $ & $ 0.49  \pm  0.02 $ & $ 62.8  \pm  6.1 $ & $ 0.11  \pm  0.01 $ & $ (3.4 \pm 0.8) \times 10^{8} $ \\ 

W3 & 87.6 & $ 0.14  \pm  0.02 $ & $ 1.15  \pm  0.02 $ & $ 62.0  \pm  6.2 $ & $ 1.14  \pm  0.02 $ & $ 62.5  \pm  6.2 $ & $ 0.09  \pm  0.01 $ & $ (7.2 \pm 1.6) \times 10^{8} $ \\ 
       
    \hline 

    \end{tabular}
    \tablefoot{(1) Component IDs. The core is in bold (2) The observed frequency (3) The flux density (4) The distance from the phase center (no core-shift applied) (5) The position angle $\phi$ (6) The distance from the core component $r_{cs}$ (7) The position angle $\phi_{cs}$ associated with $r_{cs}$ (8) The FWHM of the fitted Gaussian (9) The brightness temperature of the component.}
    \label{tab:components}
    \end{minipage}
        
\end{table}

\end{appendix}

\end{document}